\documentclass[11pt,a4paper]{article}
\pdfoutput=1

\usepackage{jcappub}
\usepackage{latexsym}
\usepackage{multirow}
\usepackage{color}
\usepackage[usenames,dvipsnames,table]{xcolor}
\usepackage{graphicx}
\usepackage{epsfig}  
\usepackage{epsf}    
\usepackage{dcolumn}
\usepackage{bm}
\usepackage{dcolumn}
\usepackage{textcomp}
\usepackage{float}
\usepackage{subfig}
\usepackage{hypcap}
\usepackage[]{hyperref}
\usepackage{makecell}
\usepackage{color}
\usepackage{pifont}
\usepackage{appendix}
  
\hypersetup{
  bookmarks=true,         
  unicode=false,          
  pdftoolbar=true,        
 pdfmenubar=true,        
 pdffitwindow=true,     
 pdfstartview={FitH},    
 pdfsubject={Neutrino Oscillations Phenomenology},   
 pdfnewwindow=true,      
 pdfcreator={RevTeX},
 colorlinks=true,       
 linkcolor=red,          
 citecolor=blue,        
 filecolor=black,      
 urlcolor=blue,           
  }

\makeatother

\title{UHECR primary identification using the lateral profile of muons in EAS}

\author[a,b,1]{Moon Moon Devi, \note{Corresponding author.}}
\author[a]{Ranny Budnik} 

\affiliation[a]{Weizmann Institute of Science, Herzl St. 234, Rehovot 7610001, Israel}
\affiliation[b]{Presently at: Tezpur University, Napaam, Assam 784028, India}

\emailAdd{devimm@tezu.ernet.in}
\emailAdd{ran.budnik@weizmann.ac.il}

\abstract
{
New developments in detector technology allow for a realistic cost of large area surface detectors for cosmic ray air showers, 
with some limitations on particle identification, energy resolutions, directional information and dynamic range. 
In this paper, we present a simulation study using CORSIKA to quantify the lateral profile of the muons at ground level, 
characterized by their energy spectrum and lateral spread, and combine it  with the depth at shower maximum (X$_{max}$) of an 
EAS initiated by a primary at energies $10^{16}$ eV -- $10^{19}$ eV. Using different primaries, we show that the combined 
muon observables and X$_{max}$ can identify the primary in a large fraction of the events, depending on the energy and the detector 
performance. This study provides important input parameters for the design of a future muon detector for surface array, which will 
be able to boost the knowledge of primaries and of the QCD interactions in the atmosphere.
}


\begin{document}
\maketitle
\flushbottom

\section{Introduction} 
\label{sect-intro}
The Ultra High Energy Cosmic Rays (UHECRs) are the most energetic particles known, and  their origin and 
composition are not well understood yet \cite{Greisen:1966jv, Zatsepin:1966jv, Hillas:2006ms, 
DeMarco:2005ia, Waxman:1999ny}. The detection of the UHECRs is difficult 
owing to the fact that their flux at higher energies is very low, and the only way to detect them is by 
observing the large cascades of secondary particles, known as the Extensive Air Showers (EAS), which are 
produced by the interaction of these cosmic rays with the atmosphere. The direct detectors placed in 
balloons or satellites, capable of probing the composition of cosmic 
primaries, reach energy up to about $10^{14.5}$ eV only \cite{Maurin:2013lwa,Boezio2003583, Alcaraz200027, 
Alcaraz2000193, 0004-637X-545-2-1135, Haino200435, Ave:2008as, Panov2014233, 1538-4357-628-1-L41, 
0004-637X-502-1-278,Aab:2014aea, Aab:2014kda, Abbasi:2015czo}. The state of the art experiments to detect the UHECRs 
are yet to be able to study all the key observables required. The energy and the incident direction of the primary 
cosmic rays into the atmosphere are reconstructed with precision, however, the composition of the primary mass 
remains an open question \cite{Aab:2016htd, ThePierreAuger:2015rha, FUJII2015418, TINYAKOV201429, Abbasi:2014sfa, Abu-Zayyad:2013qwa}. The composition at different energies 
has strong implications on  the sources, which  are not  completely known, the propagation, and possible interpretation of new 
physics involved at the highest energies. The galactic and extragalactic magnetic fields, the location 
of the source and its density evolution are still debated \cite{Abbasi:2016kgr, Abraham:2010mj, Abbasi:2009nf, Aloisio:2013hya, Liu:2013ppa}. 
Establishing the mass composition at the high and  ultra high 
energies, from the astrophysical reasons, bear fundamental importance as it would provide information on various 
origins and propagation of the cosmic rays \cite{ARNPS.285.572007, Abdo:2008aw, PhysRevLett.102.181101, PhysRevD.82.092004, 
PhysRevLett.113.221102, PhysRevLett.101.261104, PhysRevLett.106.201101, 2010ApJ718L194A, 2008Natur.456.362C}. 
A related question is the hadronic interactions at ultra high energies, beyond the reach of current accelerators, 
governing the initial evolution of the EAS. The hadronic interactions are parameterized by various models, however, 
at the highest energies all exhibit  significant uncertainties.

The muons in an Extensive Air Shower are sensitive to the primary composition and to the hadronic 
interaction properties. The muons are produced essentially from decaying charged mesons, and can be a fundamental 
tool to map the high energy hadronic interaction, as the individual particle signature is not washed out as 
in the electromagnetic component. In a typical EAS, muons represent about 10\% of all the 
charged particles. A large fraction of the muons reach the ground due to a low interaction 
cross section and a long decay time, and they have a wide lateral distribution. Most of the muons are 
produced in the upper atmosphere, typically 15 km above ground level, and lose about 2~GeV of their energy to ionization by 
the time they reach the ground. The energy and angular distribution of the muons reaching the ground 
depends on the production spectrum, their energy loss through the atmosphere and decays. For negligible 
muon decay (E$_\mu \textgreater$ 100/$\cos\theta$ GeV), negligible curvature of the Earth 
($\theta~\textless~70^{\circ}$) the overall muon number spectrum can be expressed by an approximate 
extrapolation formula, \cite{Bugaev, gaisser}

\begin{equation}
 \frac{dN_\mu}{dE_\mu d\Omega} \approx \frac{0.14~E_\mu^{-2.7}}{\rm cm^2~s~sr~GeV} \Bigg \{ \frac{1}{1+\frac{1.1~E_\mu\cos\theta}{115~\rm GeV}} + \frac{0.054}{1+\frac{1.1~E_\mu\cos\theta}{850~\rm GeV}} \Bigg \}.
 \end{equation}
Here, the two terms give the contribution of pions and charged kaons respectively. 
The contribution from charm and heavier flavors are negligible except at very high energy, and is neglected in this formula. 

The approximate expression for the total number of muons (N$_\mu$) at energies above 1 GeV, given by Greisen \cite{GreisenLateral}, is
\begin{equation}
  N_\mu ( \textgreater~ 1~ \rm GeV) \approx 0.95 \times 10^5  \bigg( \frac{N_c}{10^6}\bigg)^{3/4}, 
\end{equation}
where $N_c$ is the total number of charged particles in the shower.

The lateral spread of the muons would be larger than the other charged particles, and would depend on the factors 
like the transverse momenta of the muons at the point of their production and muon--multiple scattering. The spread 
has large fluctuation shower to shower, even showers with identical primary mass and 
energy.  The number of muons per square meter $\rho^{}_R$, as a function of the lateral distance R (m) 
from the core of the shower is expressed as 

\begin{equation}
 \rho^{}_R = \frac{dN_{\mu}}{dR^2} [R] = \frac{1.25~\rm N_\mu}{2\pi \Gamma(1.25)} \bigg( \frac{1}{320} \bigg)^{1.25} R^{-0.75} \bigg( 1+\frac{r}{320} \bigg)^{-2.5},
\end{equation}
where $\Gamma$ is the gamma function.

Large area detectors 
and appropriate discrimination against the much more abundant electromagnetic particles are needed for muon detection.
The measurement of the individual charged particles in an extensive air shower (EAS), at a surface detector array, 
would provide important distinguishing parameters to identify the cosmic primary particle. These will also 
contribute to the mapping of the very high energy interactions in the topmost layers of the atmosphere, i.e., 
beyond the reach of current accelerators, and to probe anomalies beyond QCD. The ongoing attempts to study 
individual muons are limited in their expandability to larger arrays \cite{doi:10.7566/JPSCP.9.010013, 
Aab:2017hhe}. 

New developments in detector technology 
allow for a realistic cost of large area detectors. Examples of these are gas 
filled photon sensors such as THGEMs \cite{Breskin:2010wi} or resistive plate WELL 
\cite{Rubin:2013jna,Moleri:2016hgk}, as well as other techniques developed worldwide 
such as Resistive Plate Chambers \cite{SANTONICO1981377}. The long term stability of these solutions is improving gradually, which is a major step for surface array applications. The application for a muon detector is possible through use of Ring Imaging Cerenkov, varying Cerenkov thresholds or others, and would require specific design based on the desired performance. A major part of the work performed in this paper is aimed at identifying the requirements in terms of energy, temporal and spatial resolutions that would suffice for important insights to cosmic ray physics.

This work aims at advancing towards primary identification, shower--by--shower, with 
the observables of the muon component of 
an EAS. The energy and position of the muons in a simulated EAS, combined with the depth at shower maximum, 
$X_{max}$, and the energy of the primary $E_p$, are used in a log likelihood analysis to distinguish the primaries. 
The paper is organized in the following way. Section~(\ref{sect:simulation}) describes the simulation. 
In Section~(\ref{sect:analysis}), the parametrization of the muon component of an EAS are described. 
Section~(\ref{sect:results}) discusses the results in distinguishing the showers from light and heavy primaries, 
choosing protons and Fe to be representative of the two groups. Finally, in Section~(\ref{sect:summary}), a brief 
summary of the work is presented with remarks on its possible application.

\section{Simulation details}
\label{sect:simulation}

In this section, we describe about the details of the air shower simulation. 

\subsection{EAS simulation}
\label{subsect:corsikasim}

      Here we describe the simulation of the EAS events and their translation in a horizontal array 
      of detectors. The extensive air showers, initiated by cosmic ray primaries of different masses, 
      energies and directions, are generated in CORSIKA Monte Carlo program \cite{corsika}. CORSIKA studies in detail 
      how an EAS evolves in the atmosphere, and allows to simulate the 
      interactions and decays of nuclei, electrons, photons, hadrons and muons. CORSIKA contains several 
      models and generators to treat the high energy hadronic interactions, namely the DUAL Parton Model (DPMJET)\cite{PhysRevD.51.64}, 
      the quark-gluon-string model (QGSJET01) \cite{Kalmykov:1993qe, Kalmykov:1994ys}, the mini jet model 
      SIBYLL \cite{Fletcher:1994bd, Engel:1992vf}, VENUS\cite{Werner:1993uh}, NEXUS \cite{DRESCHER200193}, EPOS LHC \cite{Pierog:2013ria}, 
      and QGSJETII-04 \cite{PhysRevD.83.014018, OSTAPCHENKO2006143}. The low energy hadronic interactions are simulated with one of FLUKA \cite{Fasso:2003xz}, GHEISHA \cite{Gheisha} 
      or UrQMD \cite{Bass:1998ca} models. The type, energy, location, direction and arrival time of all the secondary 
      particles up to thinning level are the outputs of this program. In CORSIKA, all particle decay 
      branches down to 1\% are taken into account. Thinning is not used for muons, as we follow each separately. 
      We choose a few primaries and generate showers for different primary energies and zenith angles. 
      Some of the parameters used in simulating the air showers are listed in table~\ref{Tab:corsikaparam}.  
      
\begin {table}[h]
\begin{center}
\begin{tabular}{ |l |l| }
\hline
Parameter & Value \\ 
\hline
Version & 7.4 \\  
\hline
Primary Particle & proton, Fe \\  
~ & Group I ( A $\leq$ 4 ) \\  
~ & Group II (4 $\textless$ A $\leq$ 25 ) \\  
~ & Group III (25 $\textless$ A $\leq$ 56) \\  
\hline
Zenith Angle & $0^{\circ}~\pm~2.5^{\circ}$ \\  
\hline
Azimuth Angle & $-180^{\circ}$ to $180^{\circ}$ \\  
\hline
Slope of primary energy spectrum & -2.7 \\  
\hline
Starting Altitude & 0 \\  
\hline
Observation level & 110 m above sea level \\
\hline
Earth's Magnetic field & $B_x$ = 20.40 $\mu$T \\
~ & $B_z$ = 43.23 $\mu$T \\
\hline
Hadronic interaction model ($\textless$ E$_{cm}$ = 12 GeV) & GHEISHA \\
\hline
Hadronic interaction model ($\textgreater$ E$_{cm}$ = 12 GeV) & QGSJET II--04 \\
~ & SIBYLL \\
~ & EPOS \\
\hline
 Lowest energy cut-off for hadrons (without $\pi^0$) & 0.3 GeV \\  
 Lowest energy cut-off for muons & 0.3 GeV \\  
 Lowest energy cut-off for electrons & 0.003 GeV \\  
 Lowest energy cut-off for photons (including $\pi^0$)  & 0.003 GeV \\  
 \hline
\end{tabular}
\caption{The values of some parameters used in CORSIKA--simulation}
\label{Tab:corsikaparam}
\end{center}
\end{table}
      
 \subsection{Detection}
 \label{subsect:detsim}
      
 The information (energy, direction, location) of the secondary muons produced in an EAS is then 
 translated through an array of 2 m $\times$ 2 m detectors. We 
 consider various scenarios of the separation between the detector stations, to collect 
 100\% (no separation), 1\% (20 m), 0.16\% (50 m) and 0.01\% (200 m) of the muons. 
 Apart from the case of an ideal muon detector with 100\% detection efficiency and ideal energy resolution, 
 the detector  characteristics were varied so that we analyze the prospect of a realistic detector. 
 The muon energy range used in the analysis is 0.5~GeV -- 50~GeV.

\section{The lateral spread of the muons in EAS}
\label{sect:analysis}
In this study, the lateral distribution at ground lever of the muons originating from an air shower 
is probed for different primary mass species. The difference in the average distributions 
for cosmic primaries is an important distinguishing parameter and is used to develop a statistical 
toolkit to identify the mass of the primaries.

\subsection{The average muon number density (lateral) as a function of ($E_\mu$, $R$, $X_{max}$)}

\begin{figure}[h]
\centering
\includegraphics[width=14cm]{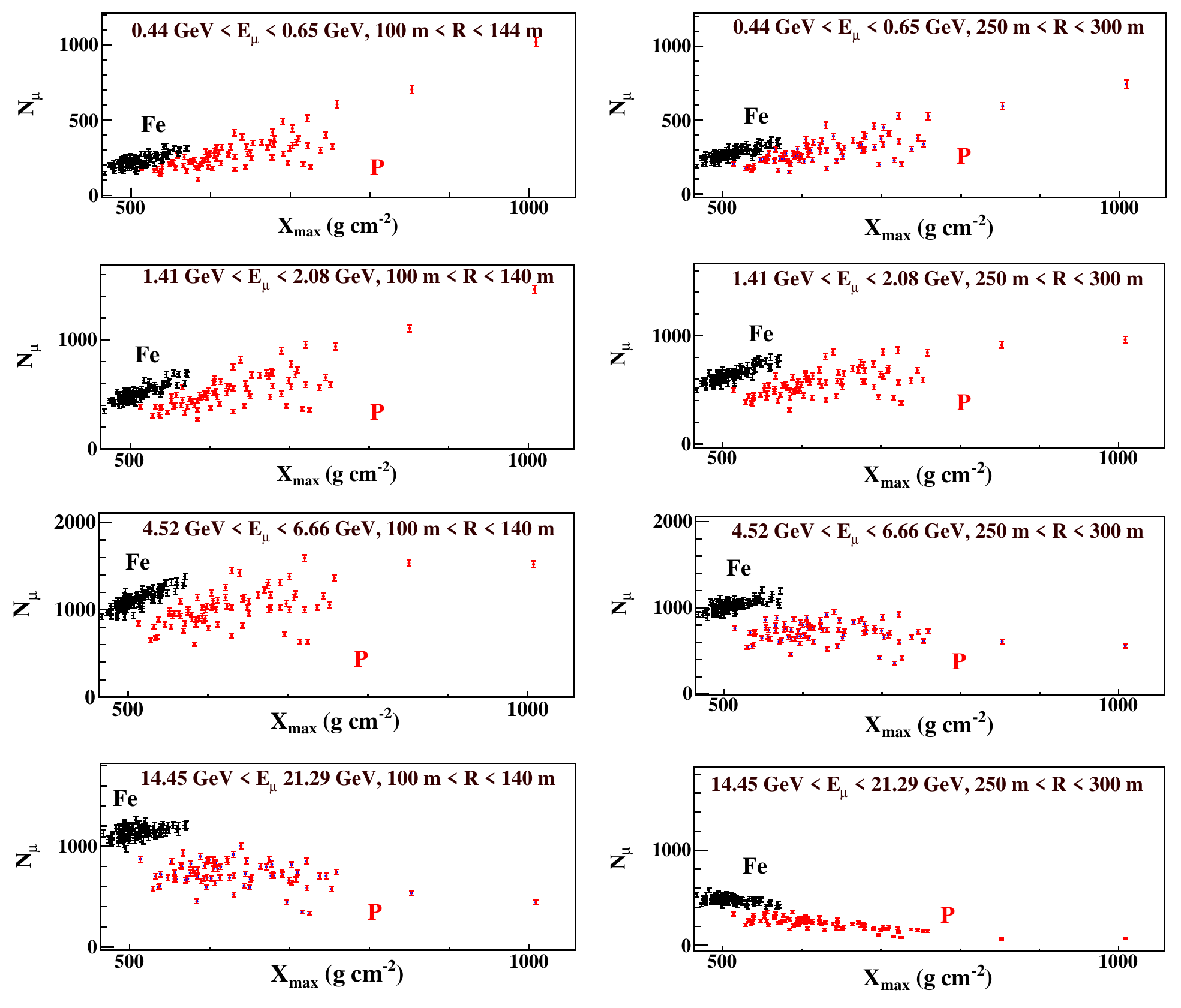}
\caption{The number of muons $N_\mu$, at a few chosen ($E_\mu$, R) bins, as a function of $X_{max}$ for 
showers with $E_p$ = 10$^{16}$ eV, $\theta_p~=~0^{\circ}~\pm~2.5^{\circ}$. The red dots represent the showers from Fe primaries, 
while the black dots represent those from proton primaries.}
\label{fig:nmu-xmax-16}
\end{figure}

We start with comparing the number of muons in proton and Fe initiated showers. In figure~\ref{fig:nmu-xmax-16}, 
we show the dependence of the number of muons ($N_\mu$) as a function of $X_{max}$, at a few chosen segments 
in $E_\mu$ and $R$. It can be seen from figure~\ref{fig:nmu-xmax-16} that, the variation in $N_\mu$ and its 
behavior with $X_{max}$ is significant even at a reasonable distance from the shower core. This is an important 
feature for this study.

\begin{figure}[h]
\includegraphics[width=6cm]{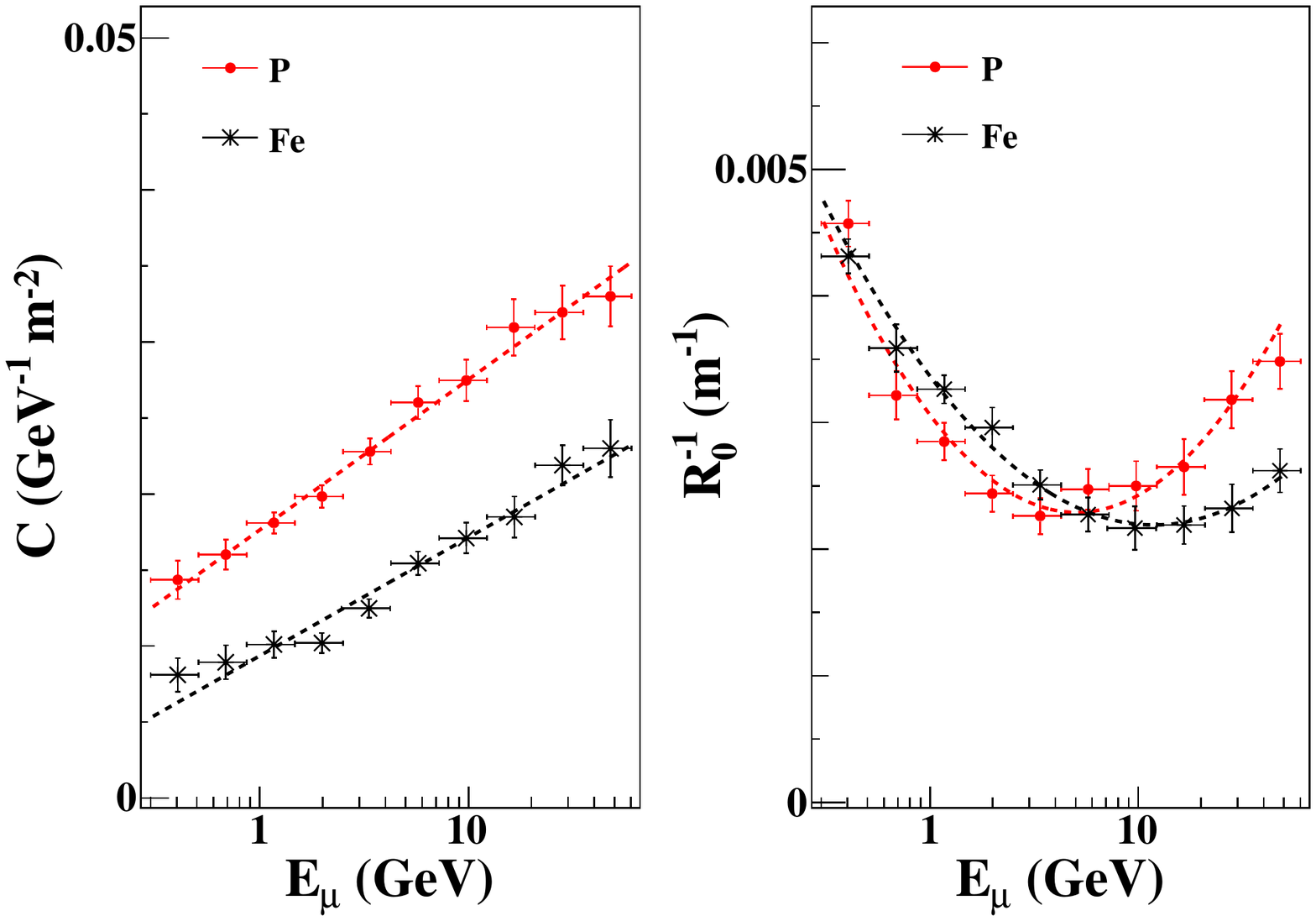}
\includegraphics[width=9cm]{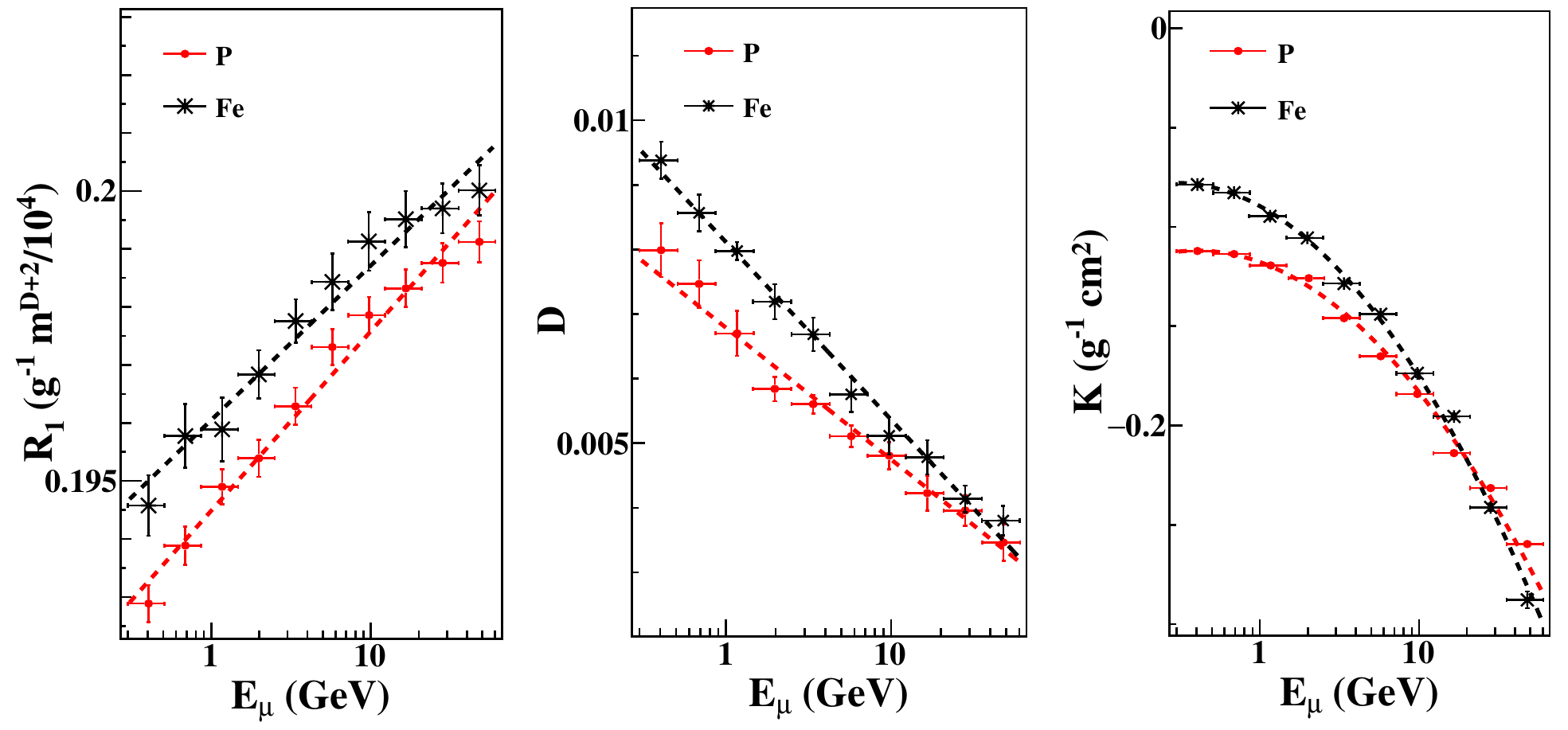}
\caption{$C$, $R_{0}^{-1}$, $R_1$, $D$ and $K$ as a function of $E_{\mu}$. $E_p$ = 10$^{16}$ eV, $\theta_p~=~0^{\circ}~\pm~2.5^{\circ}$ }
\label{fig:showerparameters}
\end{figure}

\begin{figure}[h]
\centering
\includegraphics[width=10cm]{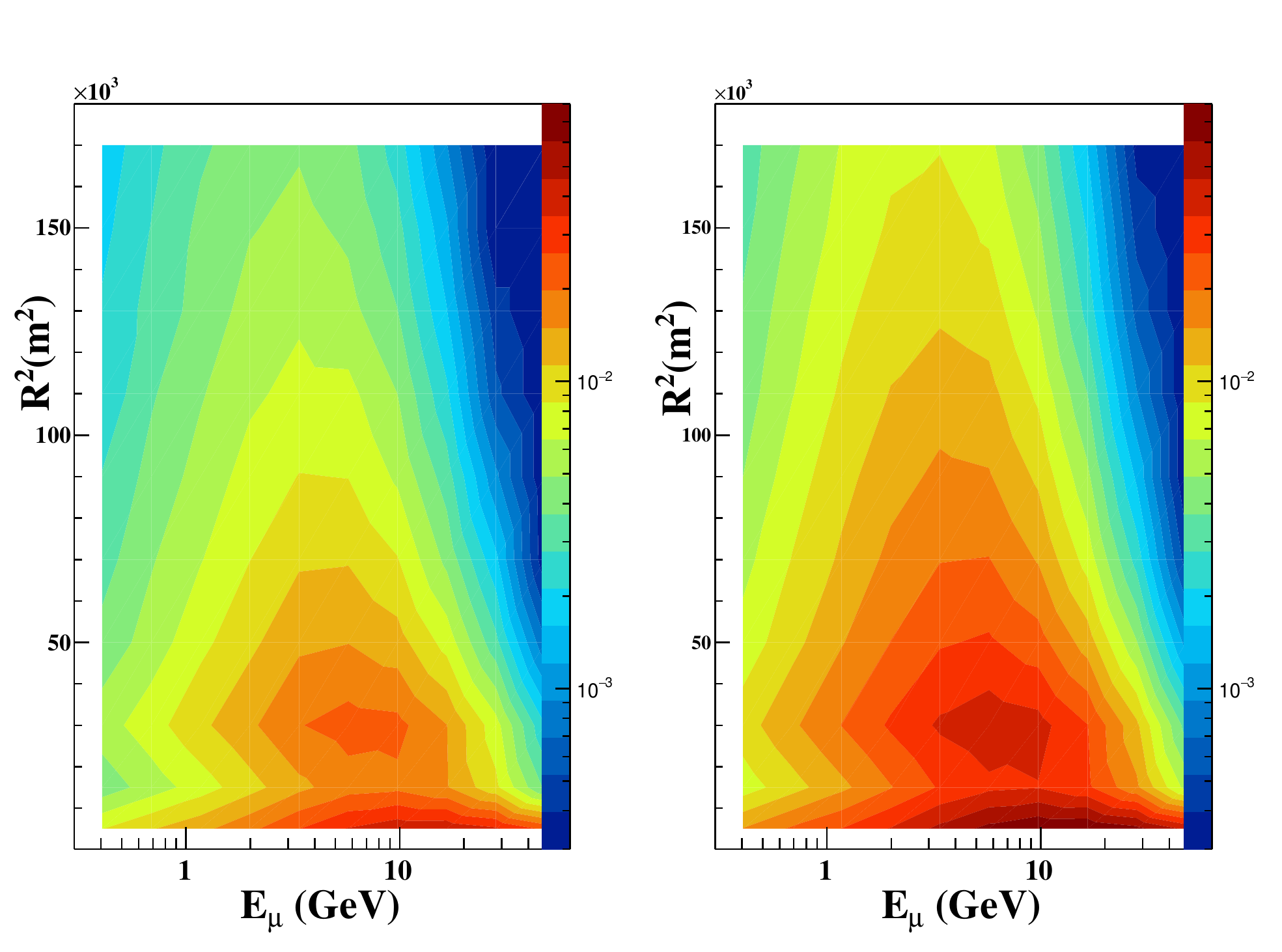}
\caption{Reconstruction of the shower profile using $\rho^{}_{ER^2}$ (GeV$^{-1}$ m$^{-2}$) for two primaries at 
$E_p~=~10^{17}$ eV, $\theta_p~=~0^{\circ}~\pm~2.5^{\circ}$. (Left) Proton initiated shower with $X_{max}$ = 621 g cm$^{-2}$, 
(Right) Fe initiated shower with $X_{max}$ = 620 g cm$^{-2}$. }
\label{fig:showerfromfunc}
\end{figure}

In this work we obtain a correlation among the muon energy, the lateral position of the 
muons with respect to the shower core, and the $X_{max}$ of the shower. 
All these parameters have significant fluctuation shower-to-shower. 
100 showers for proton and Fe each are used for best fitting.
Using the simulated events, we calculate the average lateral number density 
${\rho}^{}_{ER^2}$ (per GeV per square meter) of the muon shower as a function of $E_\mu$ (GeV),  
R (m) and $X_{max}$. We build a map in ($E_\mu$, $R$) for each primary and 
$X_{max}$ which can be expressed in a functional,

\begin{equation}
\label{eq:ro-er}
 {\rho}^{}_{ER^2} = \frac{dN_{\mu}}{dE_{\mu} dR^{2}} [ X_{max}, E_{\mu}, R] = C e^{-\frac{R}{R_0}+(R_{1} R^{-D}+K) X_{max}} =C_0^1 e^{\small -RC_2^4+(C_5^6 R^{-2C_7^8}+ C_9^{11})X_{max}},
\end{equation}

where, $C_i^j = \sum_{n=i}^{j} C_n \bar{E_\mu}^{n-i}$ and $\bar{E_\mu} = \log$ [$E_\mu$(GeV)].
Apart from the dependence on $E_\mu$, $C_i^j$s also depend on the mass, energy and direction of 
arrival of the primary cosmic ray. In figure~\ref{fig:showerparameters}, the dependence of the 
parameters C, R$_0^1$, D and K with $E_\mu$ are shown for a chosen $E_p$ and $\theta_p$.
\vspace{0.5cm}
In figure~\ref{fig:showerfromfunc}, the muon density maps of two reconstructed showers initiated by proton and Iron primaries 
respectively,  using Eqn.~(\ref{eq:ro-er}) are shown. The value of the $X_{max}$ for the two showers 
are very close to each other.

\subsection{Parametrization of the lateral muon profile for different primary mass}
\label{sect:likelihood}
The number density of muons and the lateral muon profile are then employed to track the 
primary--type using a basic likelihood test for differentiating between hypotheses. 
A likelihood function has been constructed using these two components, 

\begin{equation}
 \ln L = \ln L_{shape} + \ln L_{n},
\end{equation}
where\\
\begin{equation}
L_{shape} = \prod_{i=1}^{N_\mu^{obs}} \rho_{ER^2}^{'i} (E_\mu^i, R^i)
\end{equation}
and\\
\begin{equation}
L_{n} = Poisson(N_\mu^{obs} | N_\mu^{exp})
\end{equation}
Here, $\rho_{ER^2}^{'i}$ is the 
normalized value of $\rho_{ER^2}^{i}$ for the $i^{th}$ muon, $N_\mu^{obs}$ is the number of muons observed in an air shower, and $N_\mu^{exp}$ is the expected number of muons which is 
calculated by integrating $\rho_{ER^2}$. Now a simple likelihood using the shape profiles from the models of Proton and Fe, 
and the deviation of the data of a shower
from this average profile, gives us important discrimination between the primaries.

We define
    \begin{equation}
    \Lambda = \ln L(\rm M1) - \ln L(\rm M2)~~~~~~~~~~~~~~~~~~~~~~~~~~~~~~~~
    \end{equation}
     
   ~~~~~~~~~~~~~~~~~~~~~~~  =$\ln L_{shape} (M1) + \ln L_{n} (M1) - \ln L_{shape} (M2) - \ln L_{n} (M2)$\\
     
   ~~~~~~~~~~~~~~~~~~~~~~~  =$\ln L_{shape} (M1) - \ln L_{shape} (M2) + \ln L_{n} (M1) - \ln L_{n} (M2)$\\
     
    ~~~~~~~~~~~~~~~~~~~~~~~ = $\Lambda_{shape} + \Lambda_{n}$.\\
Here, M1 and M2 refer to two different primary types. The muon profiles $f_s$ for M1 and M2 are 
calculated using appropriate values of $C_i^j$. The number densities of the muons vary with 
different primary parameters, due to 
the difference in the interaction.
     
\section{Results}
\label{sect:results}

In this section we discuss the results of the analysis described in section~\ref{sect:likelihood}. 
We show the potential 
of the likelihood parameter $\Lambda$ in discriminating between showers from P and Fe primaries. 
Note that $\Lambda$ 
has two components $\Lambda_{shape}$ and $\Lambda_{n}$, which correspond to the contributions from the 
difference in 
the shower profile shape and the number of muons respectively. We study, with some chosen parameters of the primary 
particle, the effect of these two components for a number of detection choices.  

\subsection{P and Fe primaries}

We start with an array of ideal 2m $\times$ 2m muon detectors. In an actual surface array, only a small fraction of the 
air shower secondaries are expected to get detected. We consider four arrangements of the detector stations: the ideal 
situation of continuous detectors with no gaps between the stations, detector stations 20m apart (a collection of 1\% of the 
secondary muons), detector stations 50m apart (a collection of 0.16\% of the secondary muons) and detector 
stations 200m apart (a collection of 0.01\% of the secondary muons). The last two arrangements are close to a realistic 
array arrangements. We simulate 100 showers from each of the primary particle at a given energy. In order to have 
a significant statistics in a reasonable computation time, each of the showers is translated a number of times through the 
detector array at varying positions with respect to the shower core, and each of them are considered a new shower.
  
\begin{figure}[h]
\centering
\includegraphics[width=6cm]{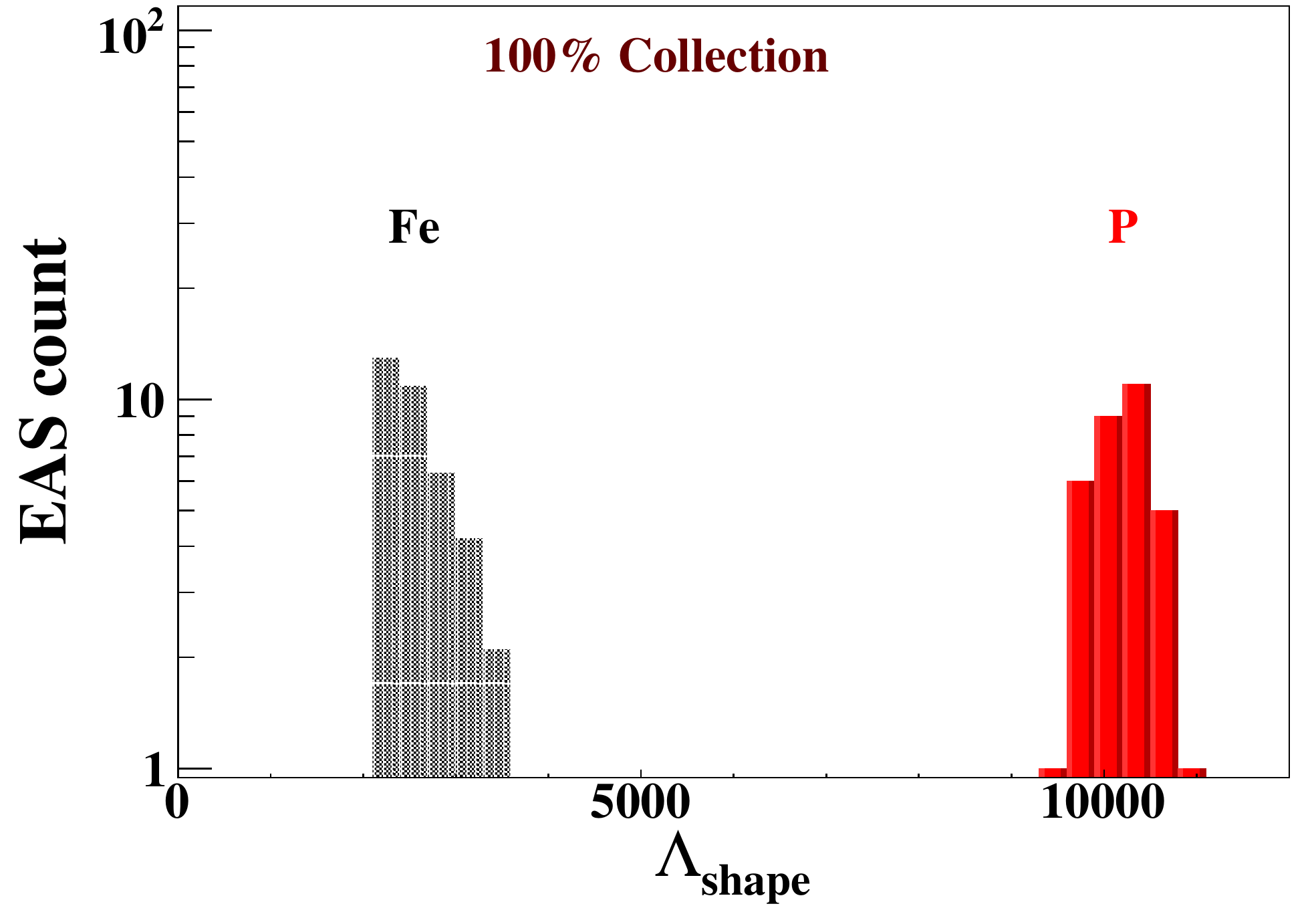}
\includegraphics[width=6cm]{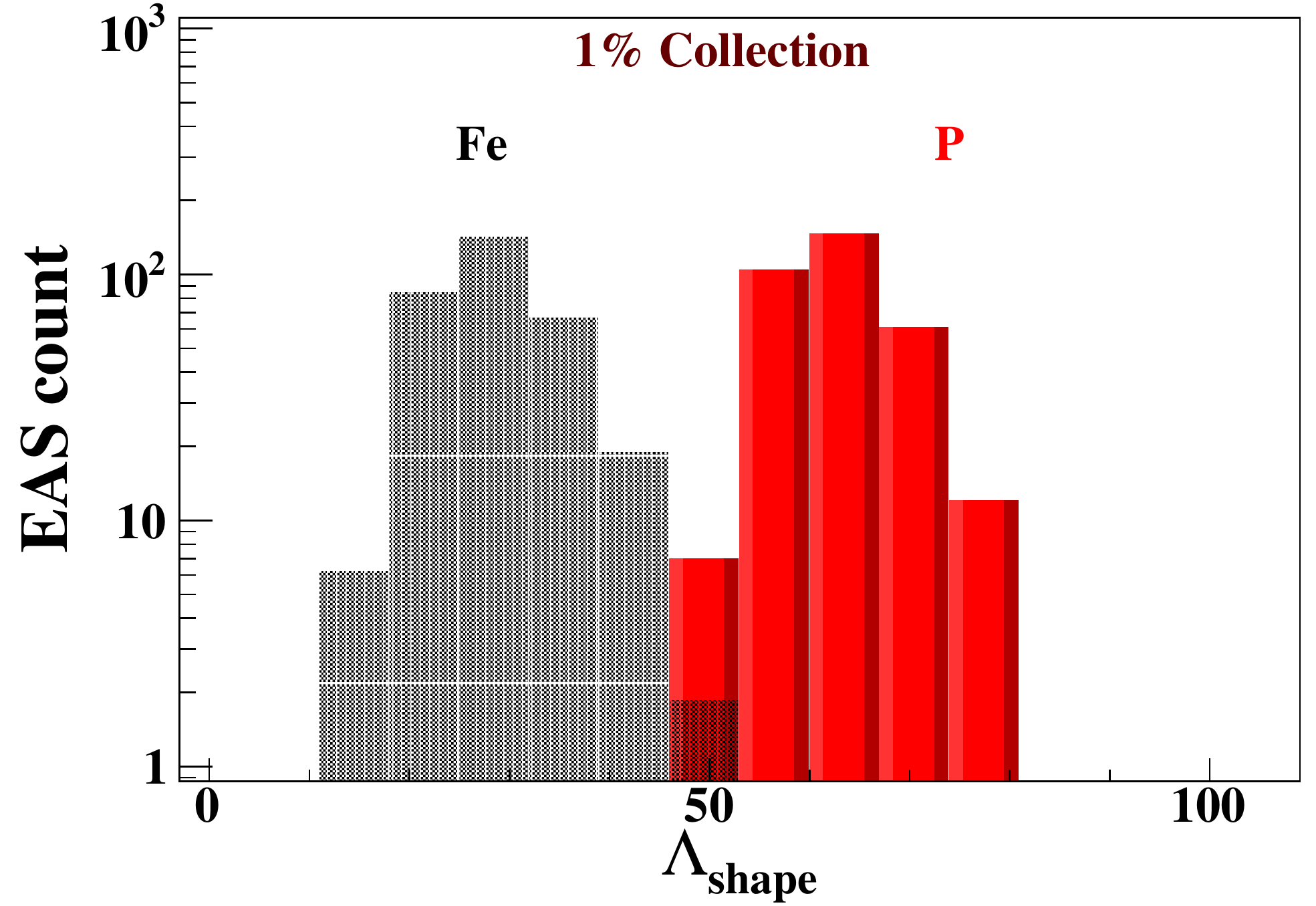}

\vspace{0.2cm}
\includegraphics[width=6cm]{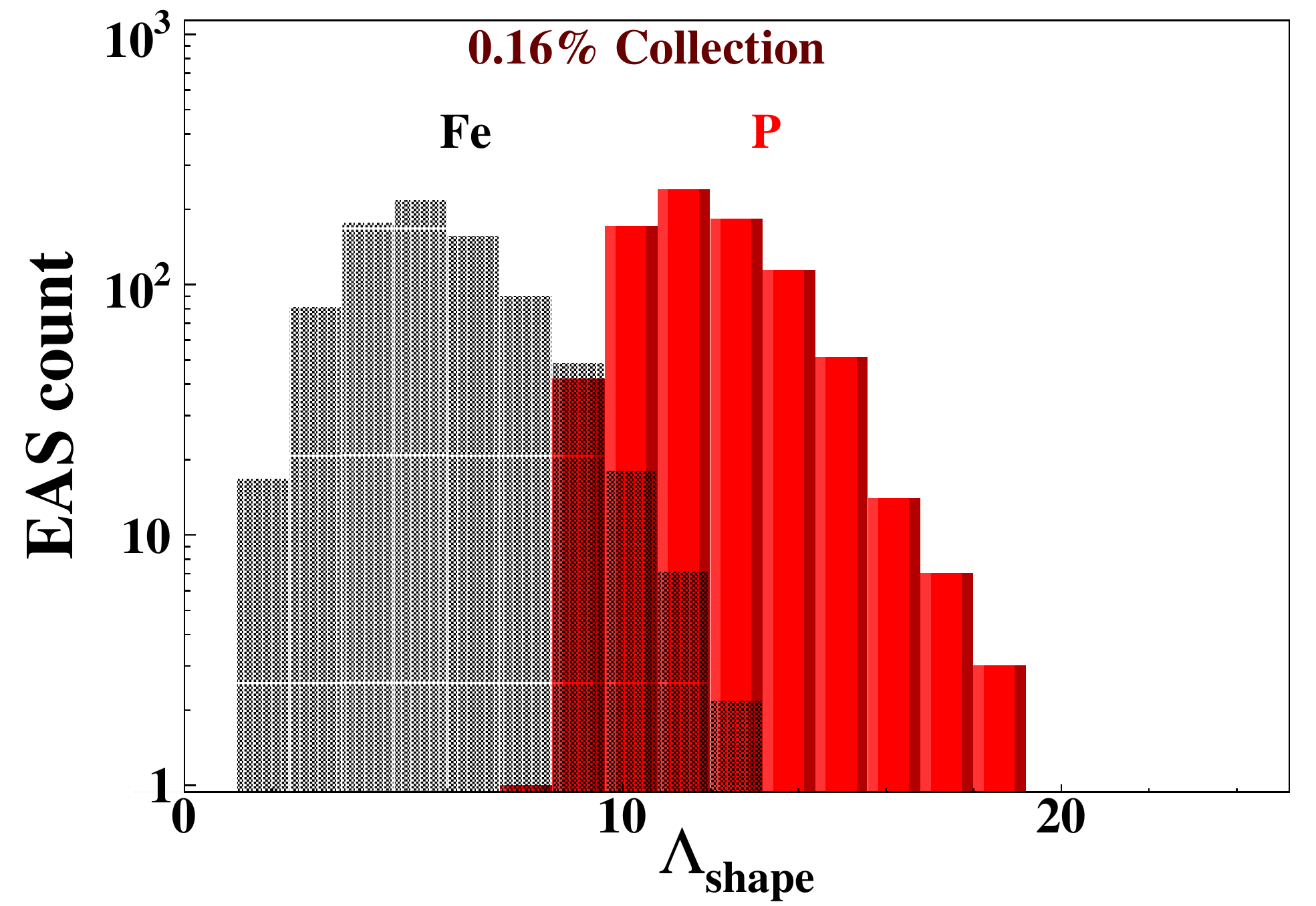}
\includegraphics[width=6cm]{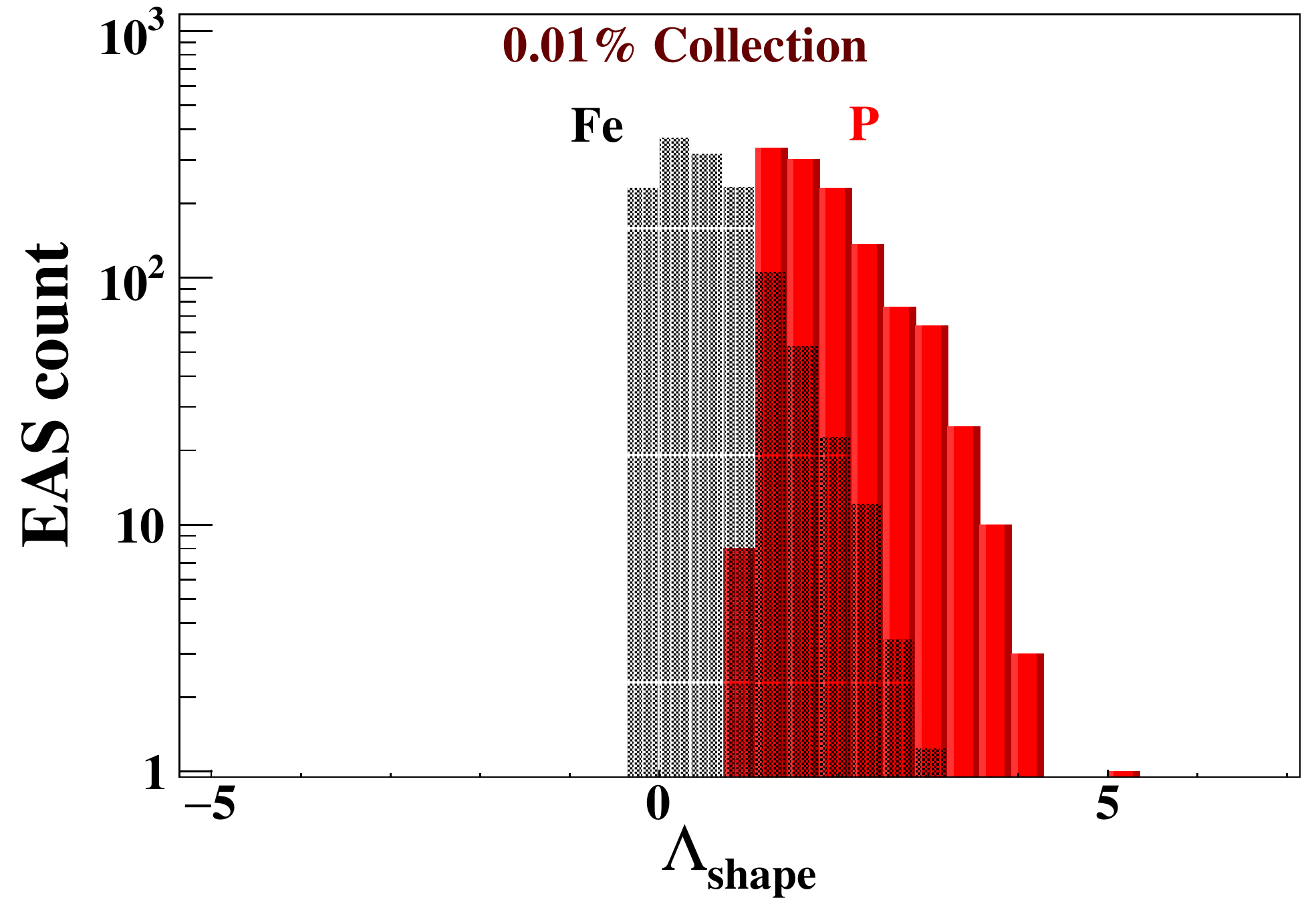}
\caption{The distributions of $\Lambda_{shape}$ for P and Fe primaries ($E_p~=~10^{16}$ eV, $\theta_p~=~0^{\circ}~\pm~2.5^{\circ}$) at four different collection efficiencies 
in an array of simulated 2m $\times$ 2m ideal muon detectors. (Top--left) 100\% collection, i.e., no gap between the detectors; 
(top--right) 1\% collection, i.e., detectors 20m apart; (bottom--left) 0.16\% collection, i.e., detectors 50m apart; 
(bottom--right) 0.01\% collection, i.e., detectors 200m apart.}
\label{fig:lambdas}
\end{figure}

  The $\Lambda_{shape}$ distributions for two sets of air showers from P and Fe primaries respectively are shown in 
  figure~\ref{fig:lambdas}. For 100\% collection efficiency the two distributions are significantly apart, raising interest 
  to probe further at realistic collections at the detector. A collection of 1\% of the entire muons with detector stations 
  20 m away from one another shows a distinction capability of more than 95\% for the two primaries. A more realistic 
  case of 0.16\% collection from a detector array with 50 m spacings indicates that the primary detection is possible with more than 
  50\% confidence. An array with 200 m spacing between stations shows rather marginal separation.

\begin{figure}[H]
\centering
\includegraphics[width=6cm]{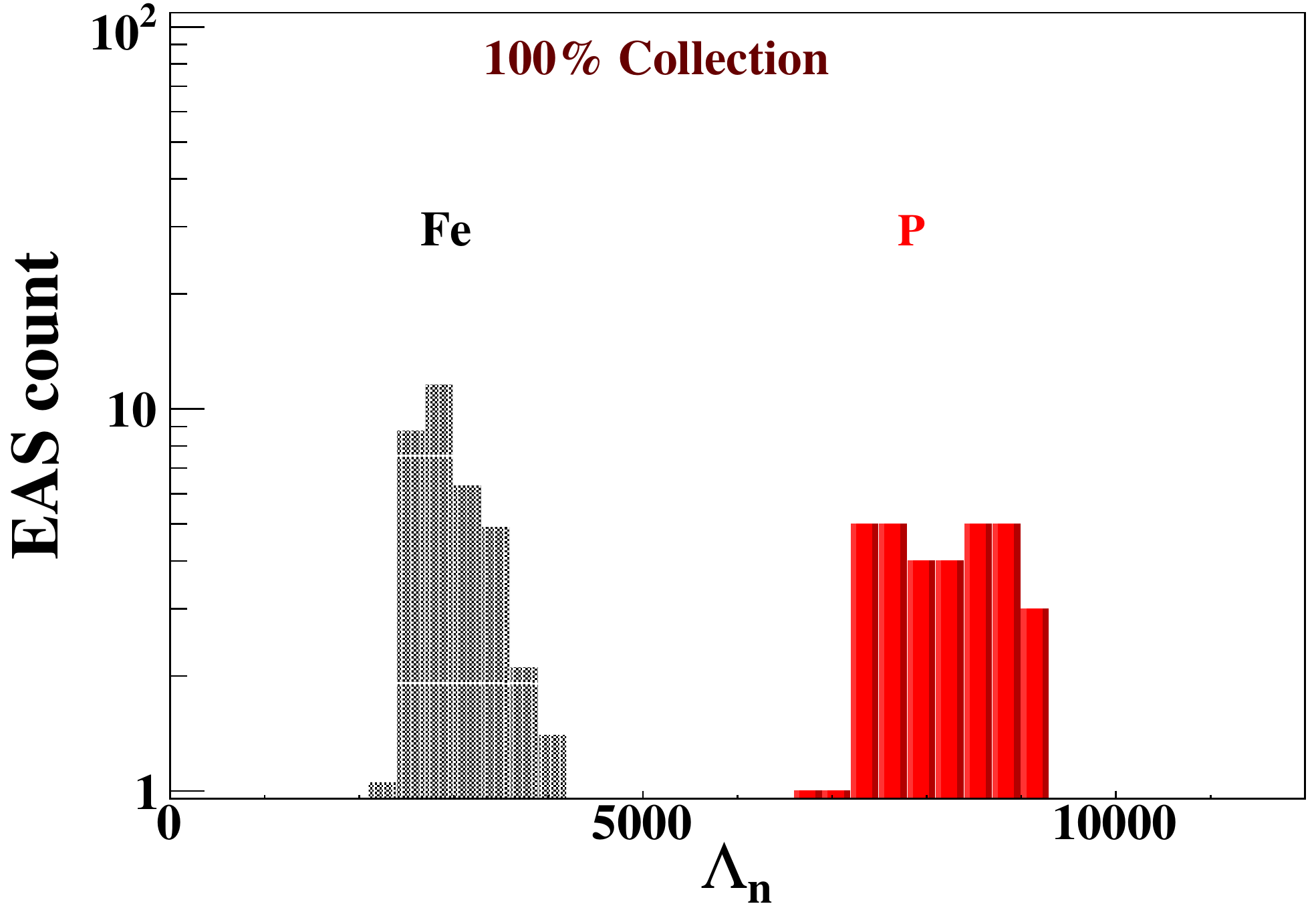}
\includegraphics[width=6cm]{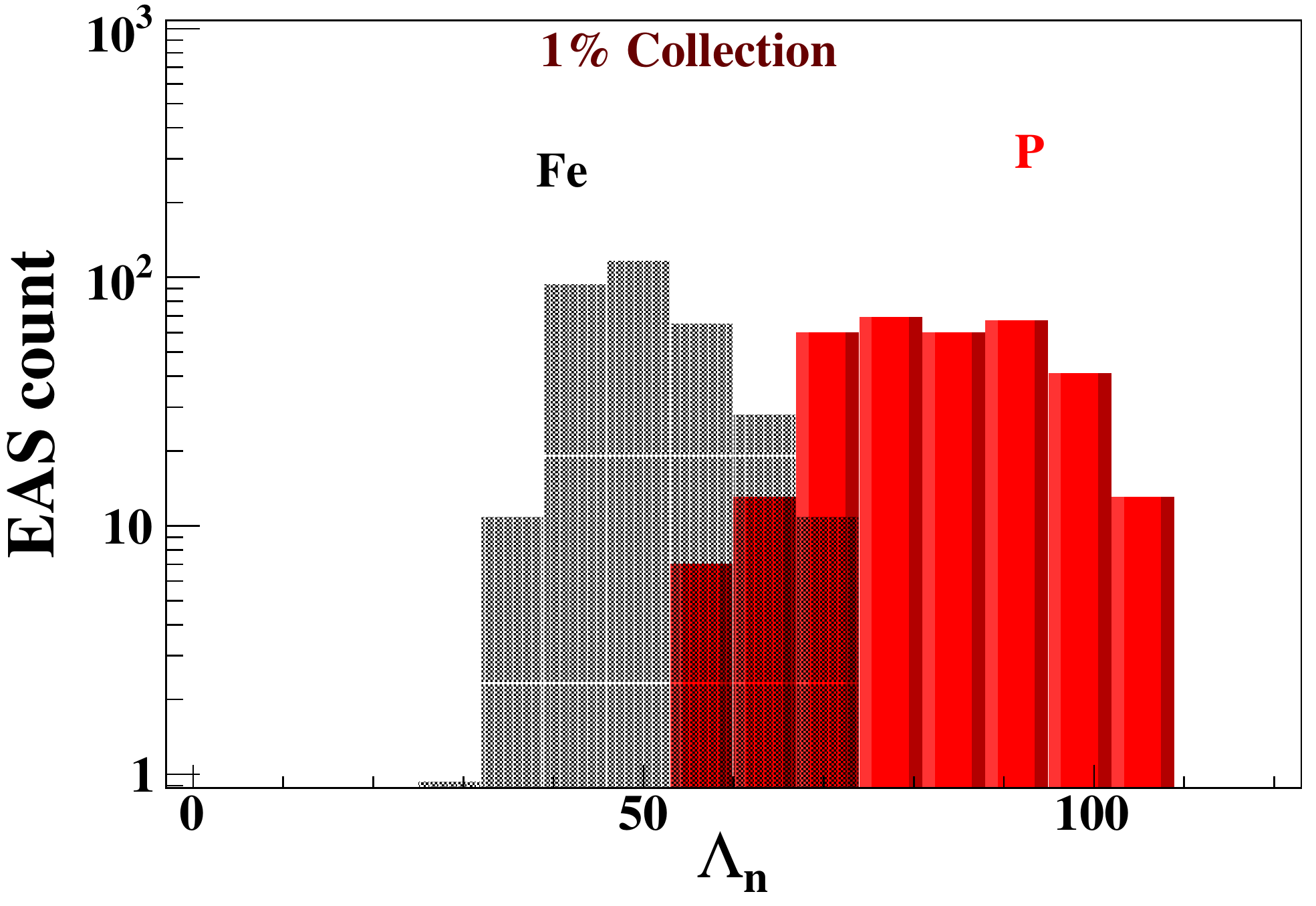}

\vspace{0.2cm}
\includegraphics[width=6cm]{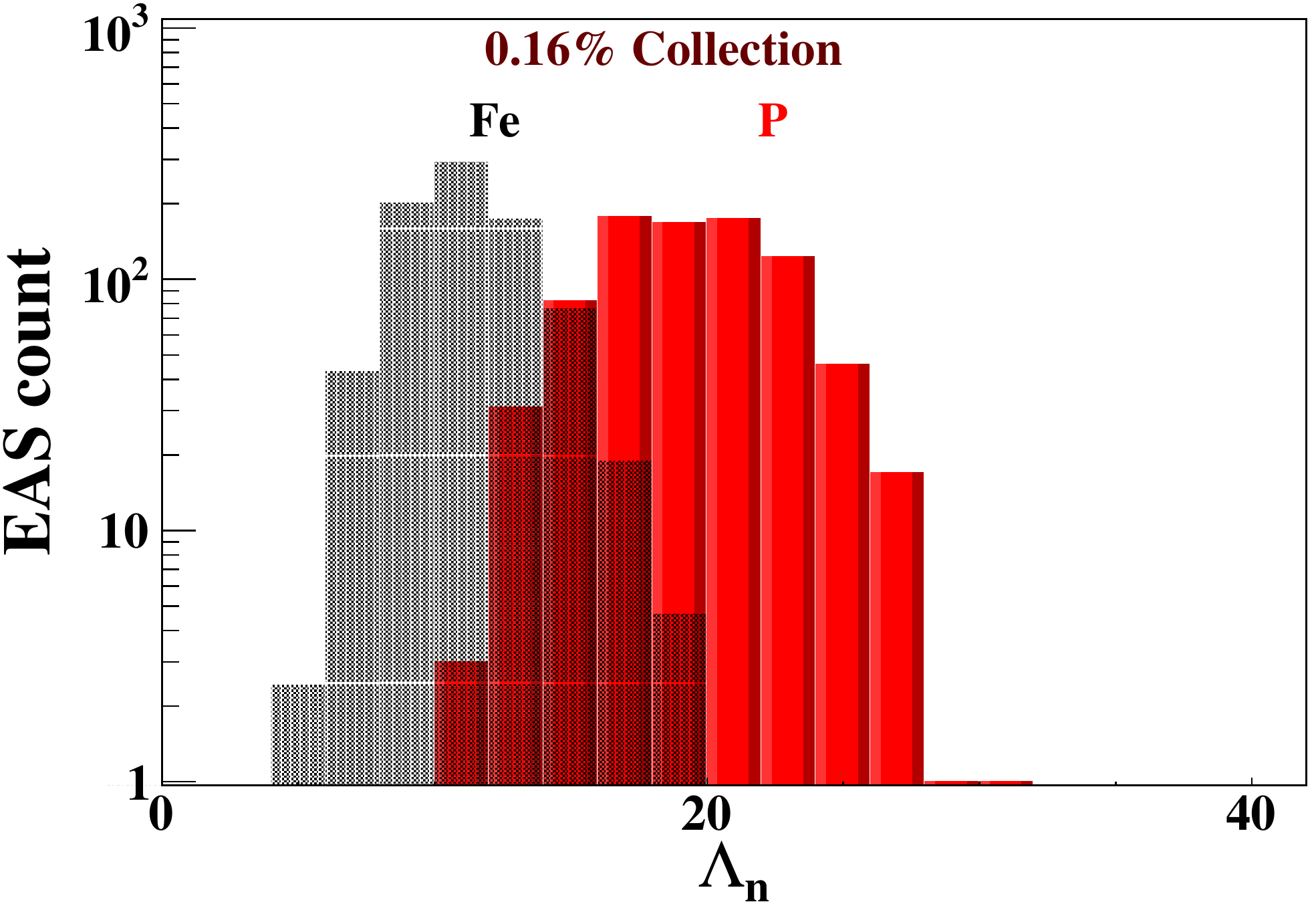}
\includegraphics[width=6cm]{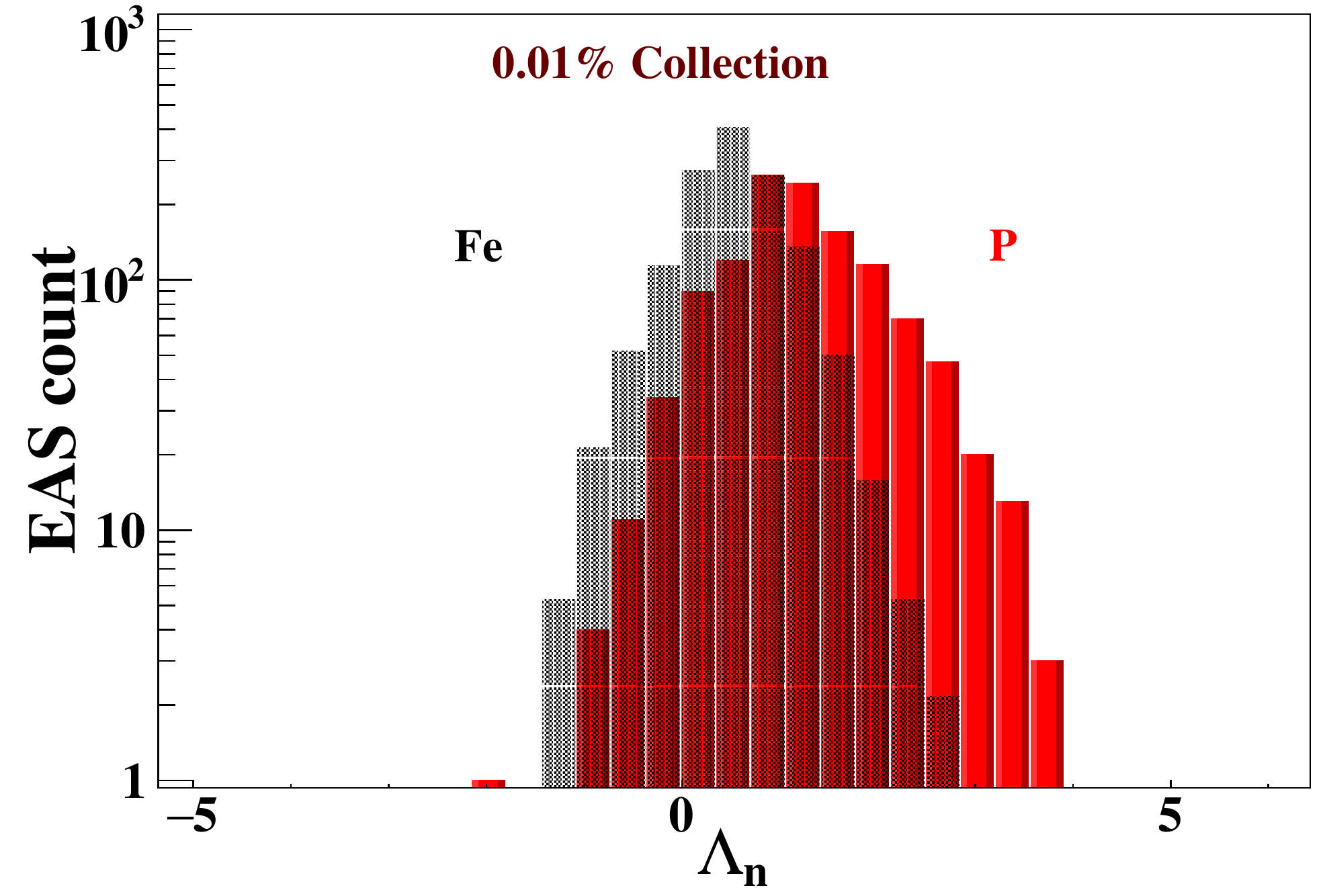}
\caption{The distributions of $\Lambda_n$ for P and Fe primaries ($E_p~=~10^{16}$ eV, $\theta_p~=~0^{\circ}~\pm~2.5^{\circ}$) at four different collection efficiencies 
in an array of simulated 2m $\times$ 2m ideal muon detectors. (Top--left) 100\% collection, i.e., no gap between the detectors; 
(Top--right) 1\% collection, i.e., detectors 20m apart; (bottom--left) 0.16\% collection, i.e., detectors 50m apart; 
(bottom--right) 0.01\% collection, i.e., detectors 200m apart.}
\label{fig:lambdan}
\end{figure}

\begin{figure}[h]
\centering
\includegraphics[width=6cm]{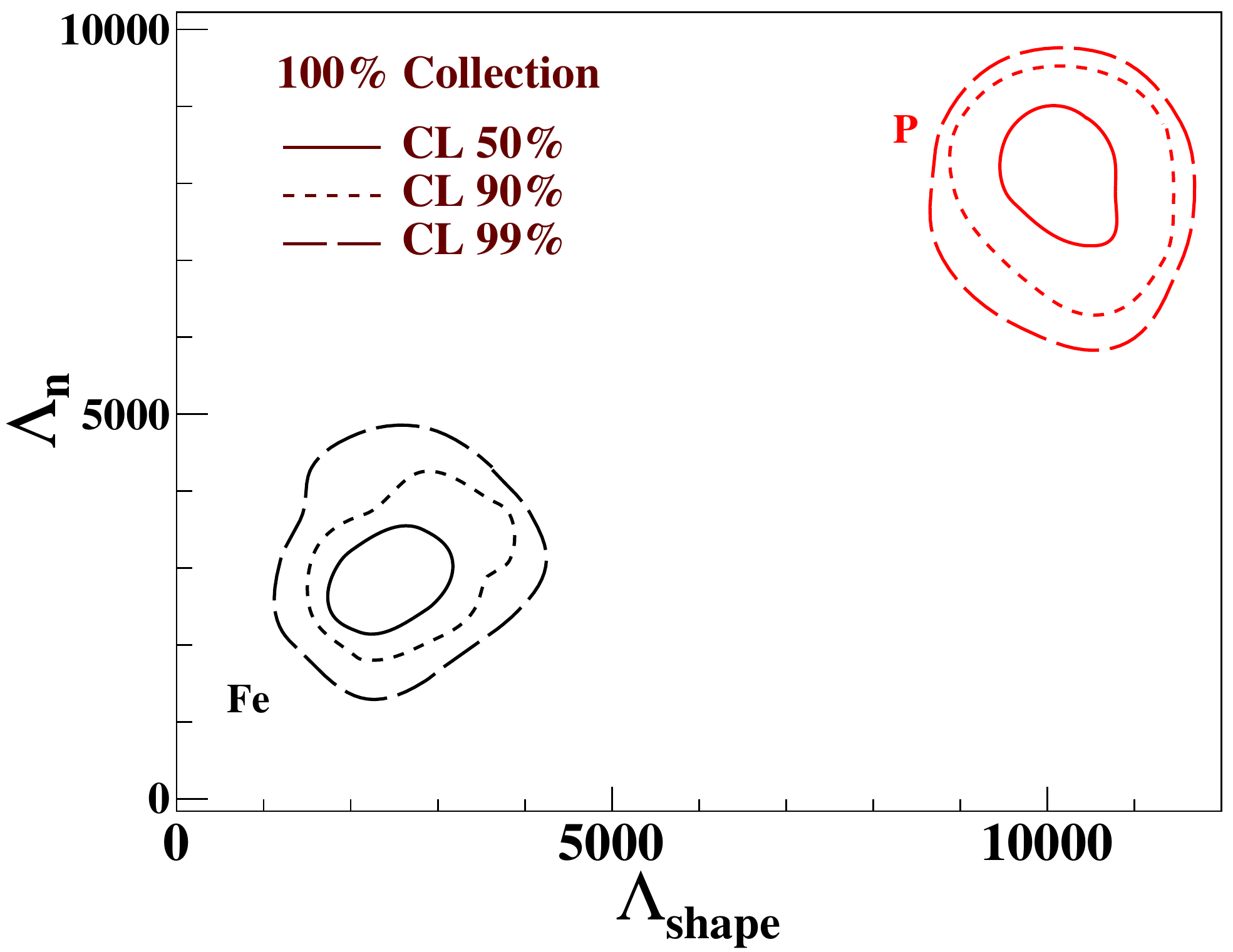}
\includegraphics[width=6cm]{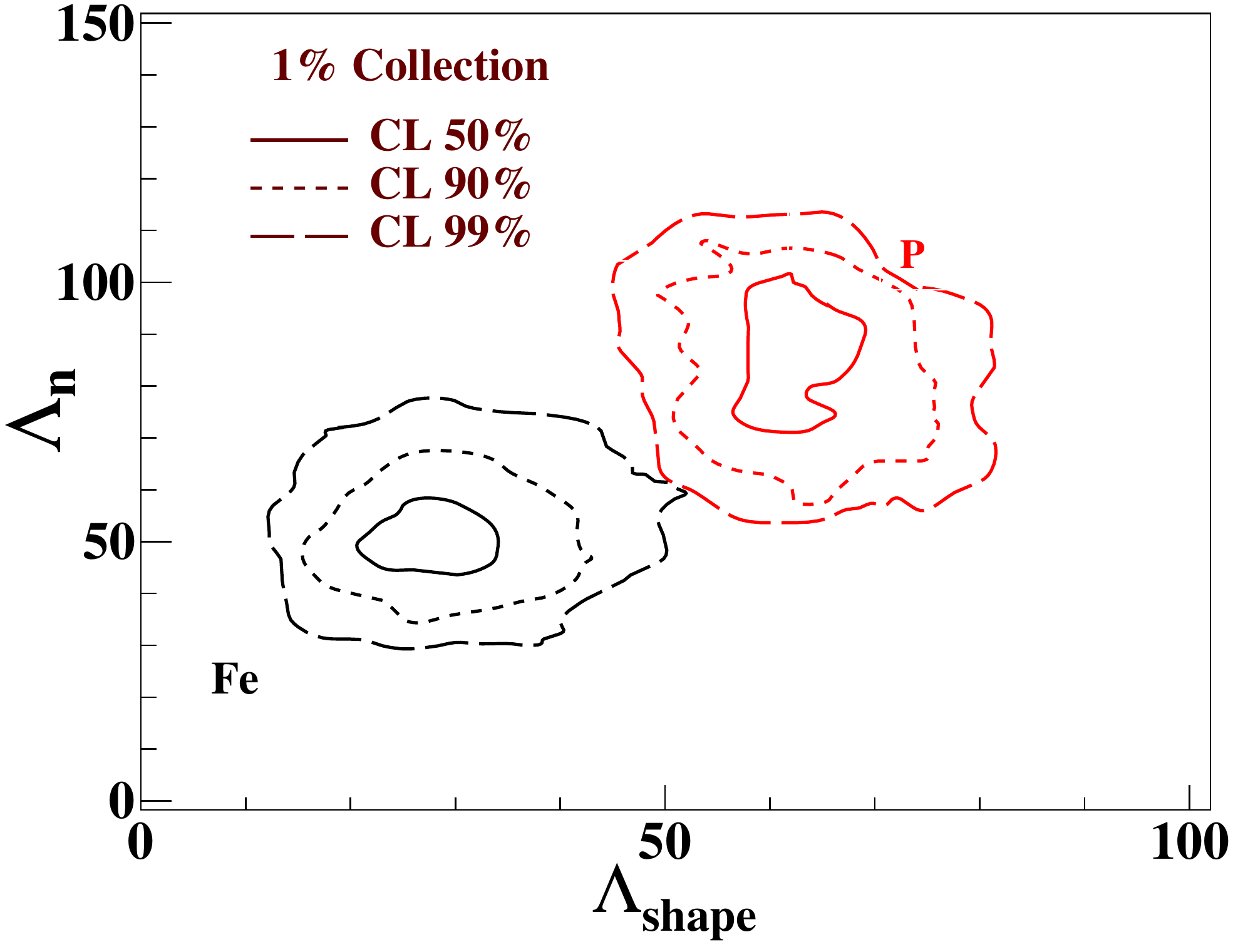}

\includegraphics[width=6cm]{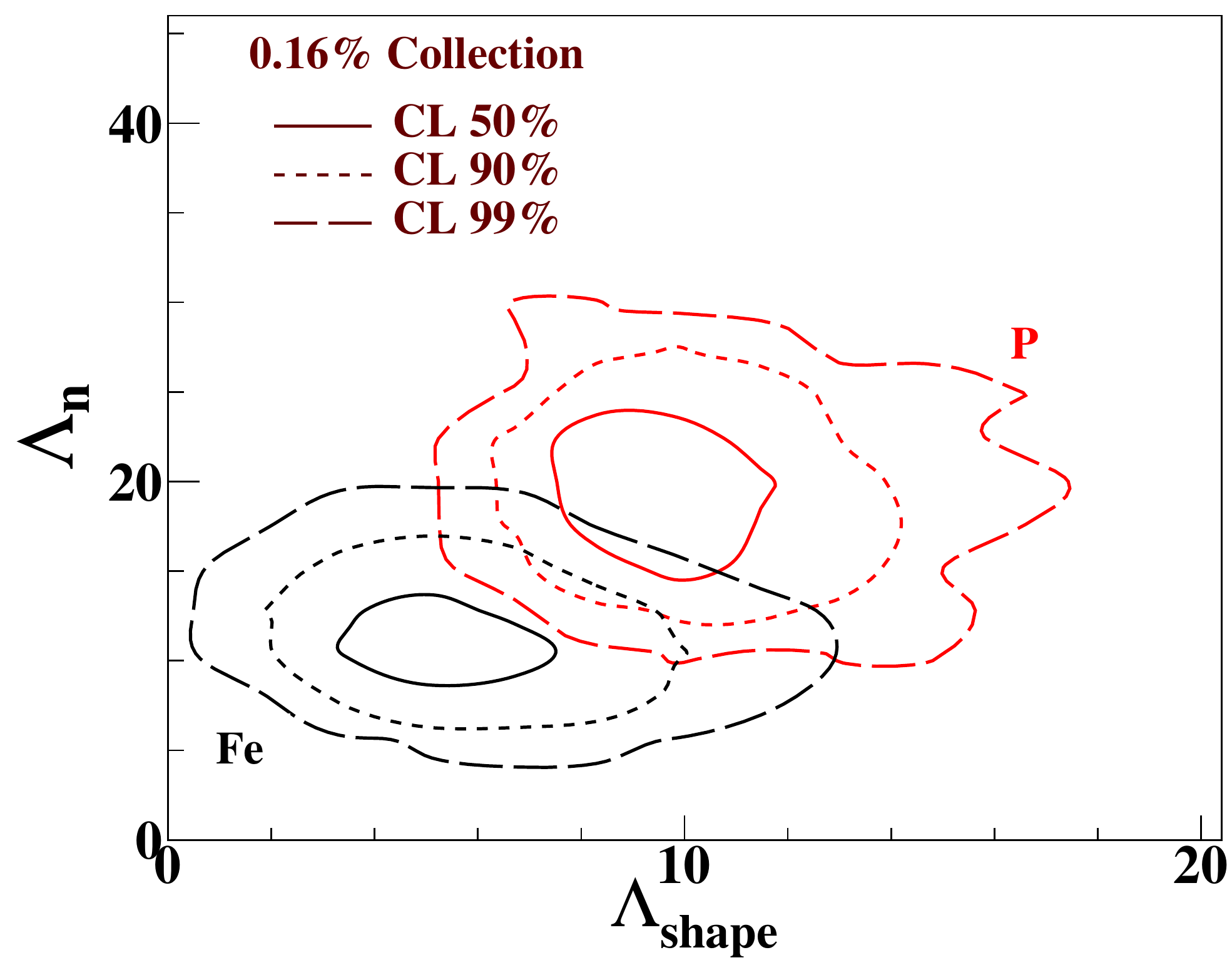}
\includegraphics[width=6cm]{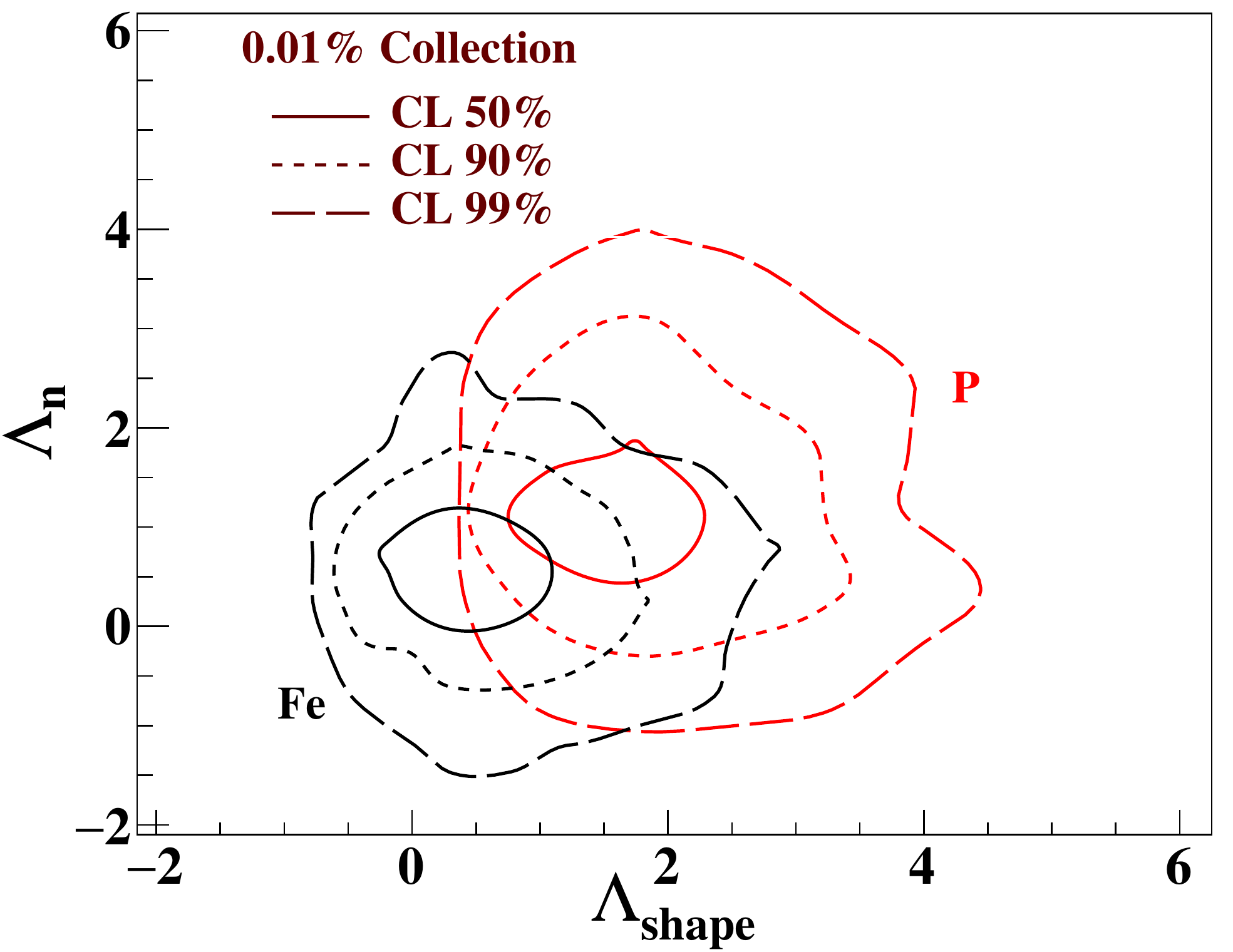}
\caption{Contours at 50\%, 90\% and 99\% confidence levels in $\Lambda_{shape}$ -- $\Lambda_{n}$ plane for P and Fe showers ($E_p~=~10^{16}$ eV, $\theta_p~=~0^{\circ}~\pm~2.5^{\circ}$)
for different collection efficiencies in an array of simulated 2m $\times$ 2m ideal muon detectors. (Top--left) 100\% collection, i.e., no gap between the detectors; 
(Top--right) 1\% collection, i.e., detectors 20m apart; (bottom--left) 0.16\% collection, i.e., detectors 50m apart; 
(bottom--right) 0.01\% collection, i.e., detectors 200m apart.
}
\label{fig:2gcevar}
\end{figure}

The distributions of $\Lambda_{n}$ for the same sets of air showers are shown in figure~\ref{fig:lambdan}. 
For the ideal 100\% collection efficiency, the two distributions are again wide apart. For detector stations 
20 m apart (1\% collection), the two primaries can be identified with about 75\% confidence. For 0.16\% collection, 
the separation capability is slightly less than 50\%. For 0.1\% collection, the identification capability is marginal. 
The two parameters ($\Lambda_{shape}$ and $\Lambda_{n}$) together will enhance the primary identification capability. 
In figure~\ref{fig:2gcevar} the projection of the parameters on the two dimensional $\Lambda_{shape}$ -- $\Lambda_{n}$ 
plane are shown. Confidence level contours at three different confidence levels (50\%, 90\% and 99\%) are drawn for 
air showers from both P and Fe primaries.

\begin{figure}[h]
\centering
\includegraphics[width=14cm]{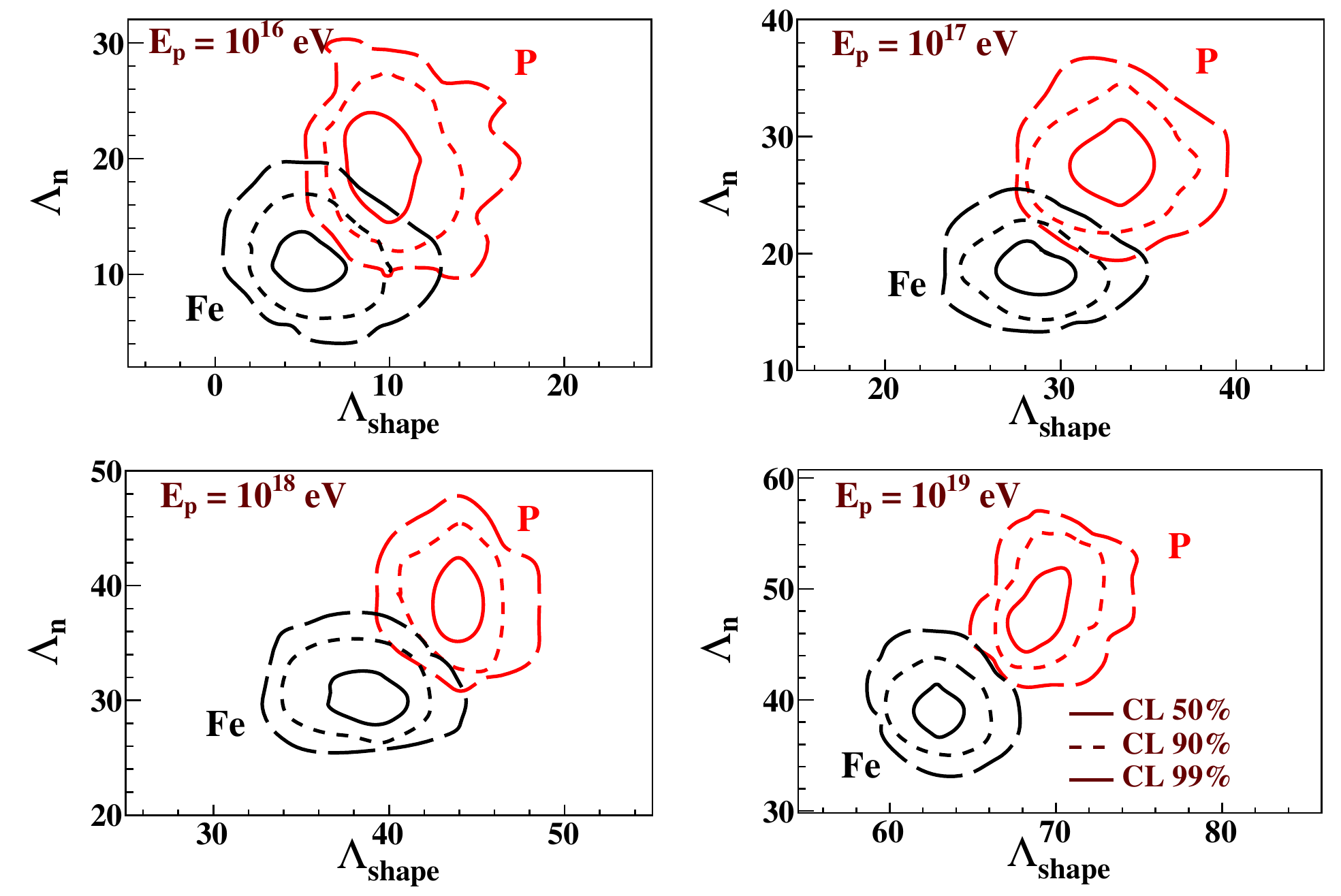}

\caption{Contours at 50\%, 90\% and 99\% confidence levels in $\Lambda_{shape}$ -- $\Lambda_{n}$ 
plane for P and Fe showers ($\theta_p~=~0^{\circ}~\pm~2.5^{\circ}$) at various energies of the primary. 
(Top--right) $E_p~=~10^{16}$ eV; (top--left) $E_p~=~10^{17}$ e; (bottom--right) $E_p~=~10^{18}$ eV, 
(bottom--left) $E_p~=~10^{19}$ eV. An array of 
2m $\times$ 2m ideal muon detectors is considered with collection efficiency 0.16\%.}
\label{fig:2gp16ceEvar}
\end{figure}

In figure.~\ref{fig:2gp16ceEvar}, we show the confidence level contours for air showers generated from 
different primary energies. A rectangular array of 2m $\times$ 2m ideal muon detector stations situated 
50 m apart from each other, which implies a collection efficiency of 0.16\% is considered here. 
As the primary energy increases, the number of muons increases and the shape of the muon distributions 
on $E_\mu$ -- R plane can be expressed through the density functions $\rho_{ER}$ with lesser uncertainty. 
This improves the identification of the primary mass through $\Lambda_{shape}$ and $\Lambda_{n}$, which 
is clear from figure.~\ref{fig:2gp16ceEvar}.

\newpage
\subsection{Incorporating $E_\mu$ resolution}

\begin{figure}[h]
\centering
\includegraphics[width=12cm]{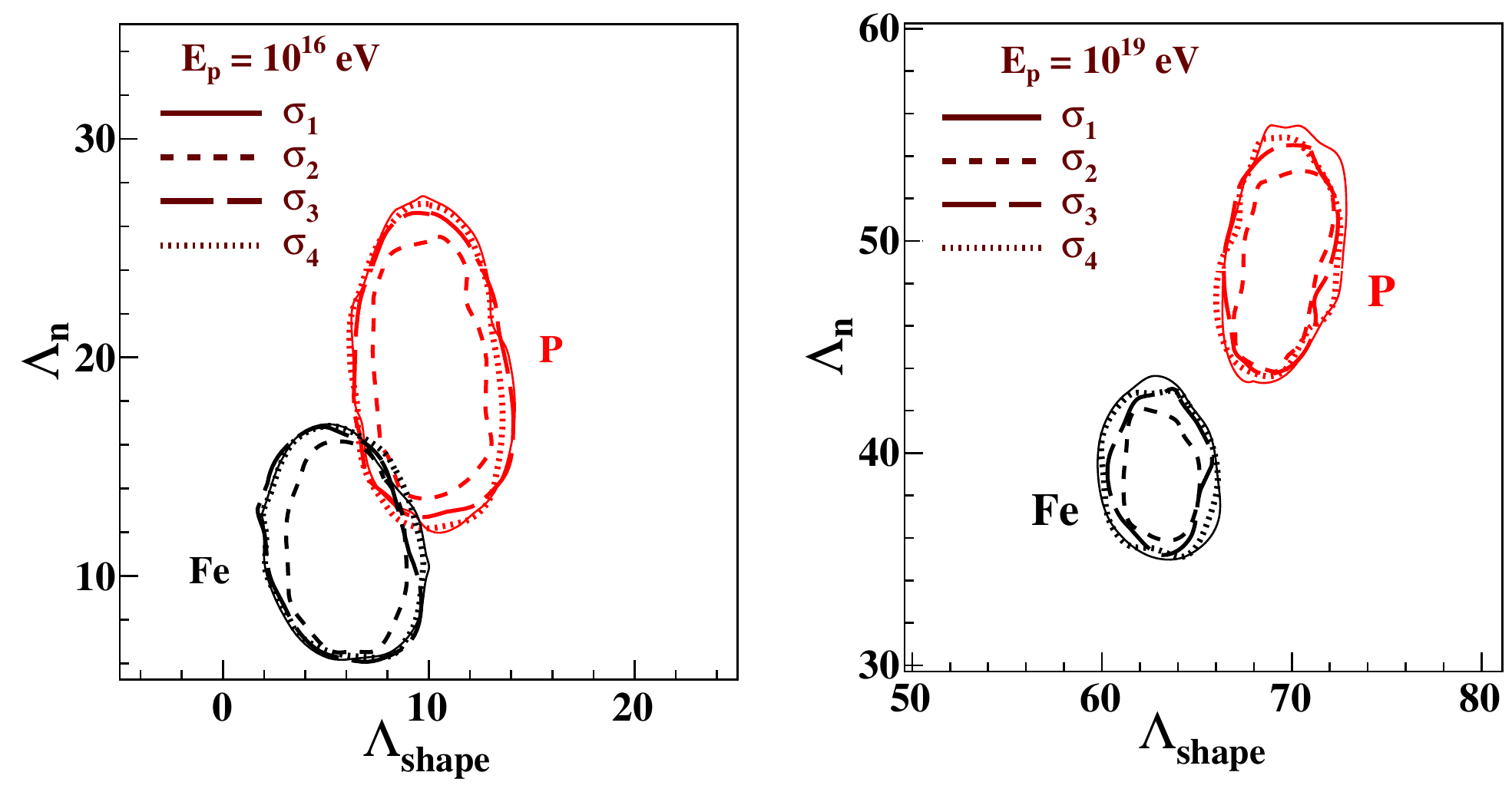}

\caption{Contours at 90\% confidence level in $\Lambda_{shape}$ -- $\Lambda_{n}$ plane for P and Fe primaries ($\theta_p~=~0^{\circ}~\pm~2.5^{\circ}$), at four 
different choices of $\sigma_{E_\mu}$ of the detectors, as detailed in table~(\ref{Table:Emures}). (Left) $E_p$ = $10^{16} eV$, (right) $E_p$ = $10^{19} eV$. 
An array of 2m $\times$ 2m muon detectors is considered with collection efficiency 0.16\%}
\label{fig:EmuRes}
\end{figure}

All the results so far have been obtained for ideal muon detectors. We now add realistic detector 
resolution to $E_\mu$ which is crucial to design a suitable muon detector. Figure~\ref{fig:EmuRes} 
shows the 90\% contours for four choices of the $E_{\mu}$ resolutions as listed in table~\ref{Table:Emures}, 
which are added as Gaussian smearing to the true values of $E_{\mu}$. The choice $\sigma_2$, 
which includes a finer resolution for $E_\mu~\leq$ 10 GeV, shows marginal improvement. 
At the detector surface level, 
the muon energy spectrum peaks at a lower energy (\textless 2 GeV) indicating that the low energy 
region is the most sensitive region for the analysis with muon number and lateral shape. 

\begin {table}[h]
\begin{center}
\begin{tabular}{ |l |l| }
\hline
$\sigma_{1}$ & 50\% \\ 
\hline
$\sigma_{2}$ & 20\%, $E_\mu~\leq$ 10 GeV \\ 
~& 50\%, $E_\mu~\textless$ 10 GeV\\
\hline
$\sigma_{3}$ & 20\%, 10 Gev $\leq E_\mu~\leq$ 20 GeV \\ 
~& 50\%, $E_\mu~\textless$ 10 GeV \& $E_\mu~\textgreater$ 20 GeV \\
\hline
$\sigma_{4}$ & 20\%, $E_\mu~\geq$ 20 GeV \\ 
~& 50\%, $E_\mu~\textless$ 20 GeV\\
\hline
\end{tabular}
\caption{The choices of $\sigma_{E_\mu}$}
\label{Table:Emures}
\end{center}
\end{table}

\subsection{More primary species}
\begin{figure}[h]
\centering
\includegraphics[width=8cm]{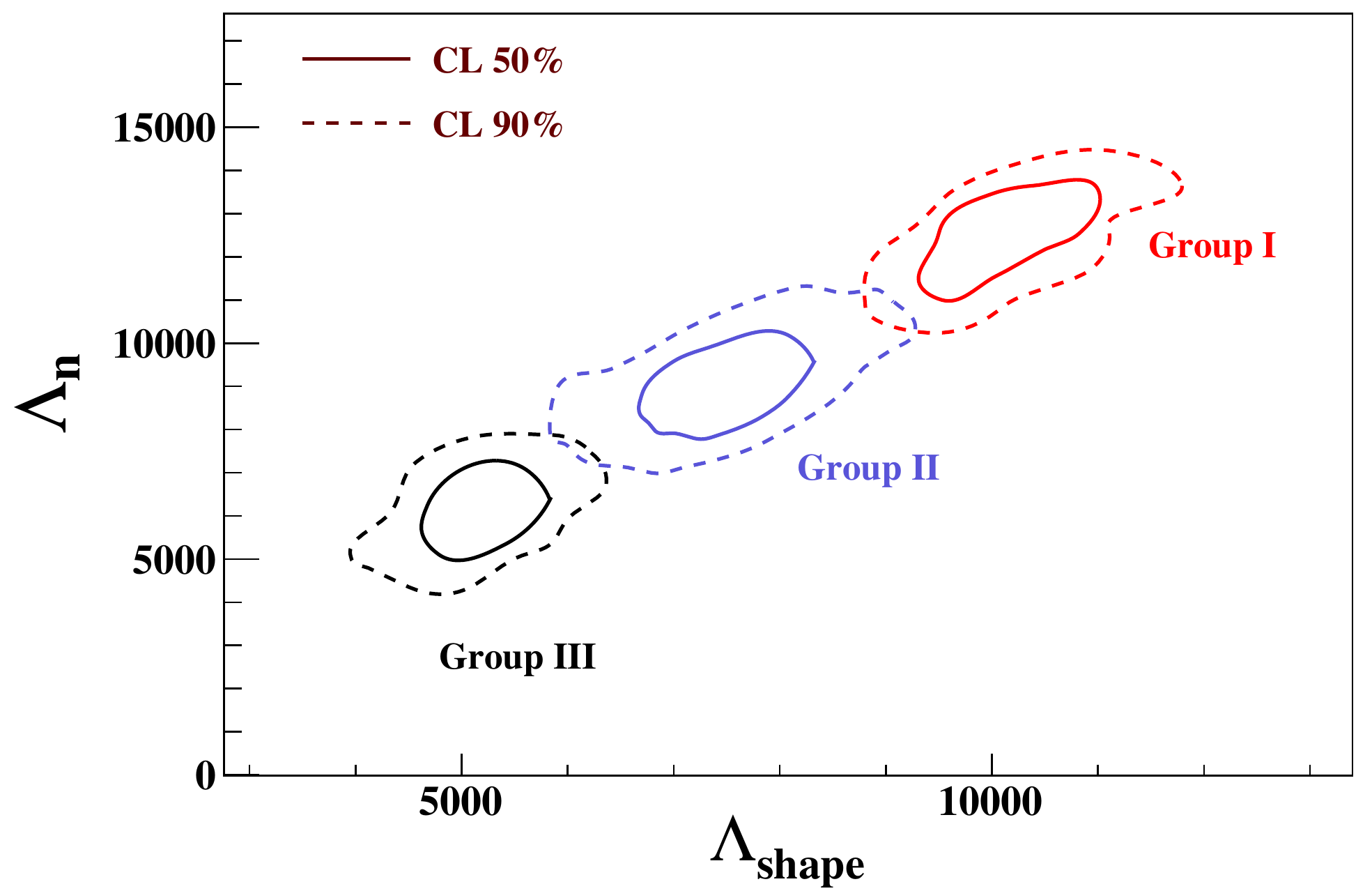}
\caption{Contours at 50\% and 90\% confidence levels in $\Lambda_{shape}$ -- $\Lambda_{n}$ 
plane for three groups of primary chosen on the basis of primary mass. Here, $E_p~=~10^{16}$ eV and 
$\theta_p~=~0^{\circ}~\pm~2.5^{\circ}$. An array of 
2m $\times$ 2m ideal muon detectors is considered with collection efficiency 100\%. }
\label{fig:3g100ce}
\end{figure}

We now extend the analysis to a wide range of primary species.
The air showers under study are distributed into three groups according to 
the mass number of their primaries as shown in Table~\ref{Tab:corsikaparam}. 
We use 200 air showers in each group, distributed evenly among the primaries.
We denote $M_{I}$, $M_{II}$ and $M_{III}$ to be the corresponding shower profile 
models of Group I, Group II and Group III respectively. We obtain the following 
distinguishing parameters from the log--likelihood study.
\begin{equation}
\Lambda(k) = \ln~L(M_{k})~-~\ln~L(M_{III});~\rm{where}~k= I, II, III
\end{equation}
In figure~\ref{fig:3g100ce}, the $\Lambda_{shape}$ -- $\Lambda_{n}$ 2-dimensional contours are drawn 
for the three groups for the most ideal case of 100\% collection with ideal muon detectors. 

\begin{figure}[h]
\centering
\includegraphics[width=12cm]{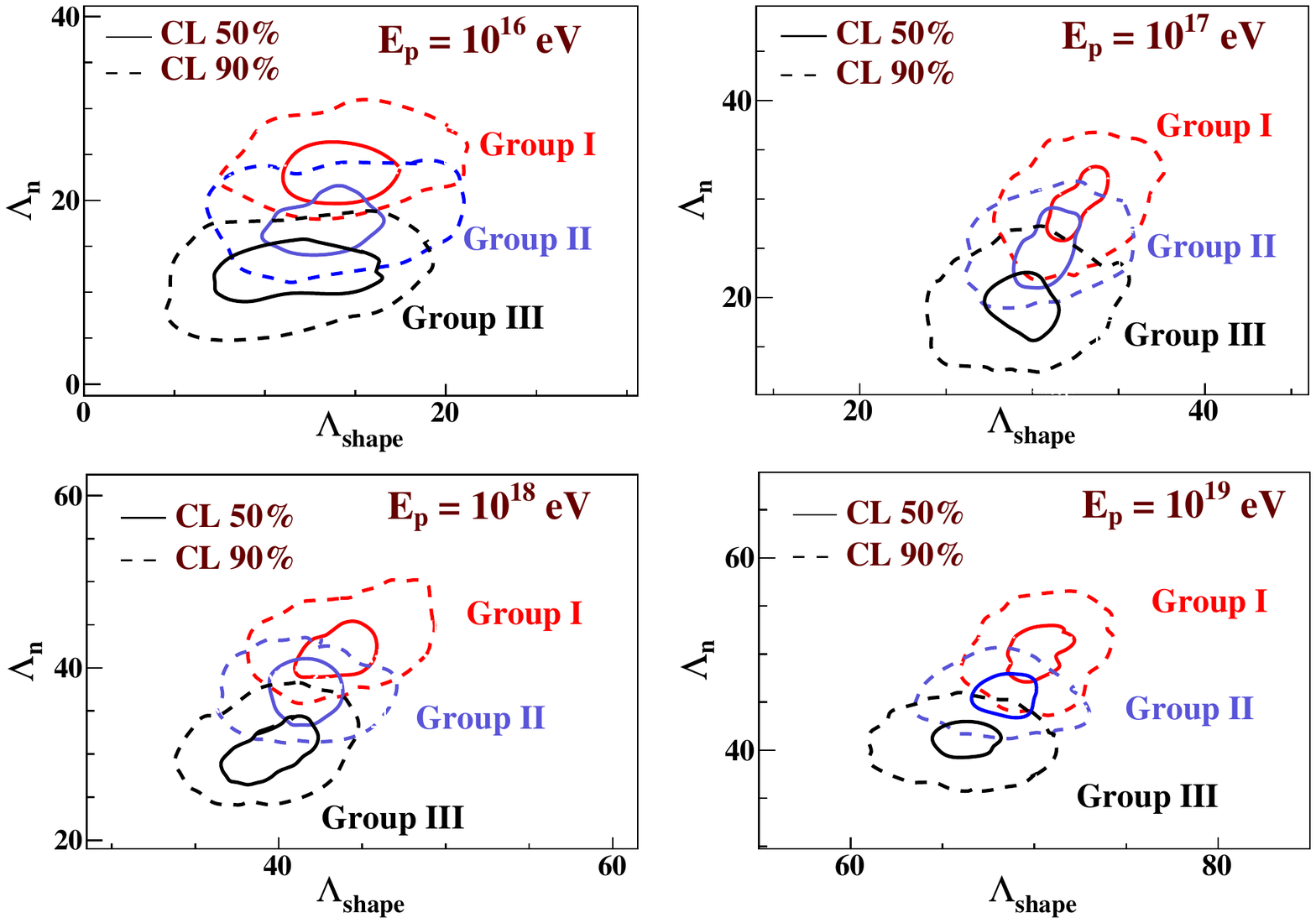}

\caption{Contours at 50\% and 90\% confidence levels in $\Lambda_{shape}$ -- $\Lambda_{n}$ plane for 
three groups of primary ($\theta_p~=~0^{\circ}~\pm~2.5^{\circ}$) chosen on the basis of primary mass.  (Top--right) $E_p~=~10^{16}$ eV; (top--left) $E_p~=~10^{17}$ eV; 
(bottom--right) $E_p~=~10^{18}$ eV, (bottom--left) $E_p~=~10^{19}$ eV. An array of 
2m $\times$ 2m muon detectors ($\sigma_{E_\mu}$ = 20\% for $E_\mu~\textless~10$ GeV, $\sigma_{E_\mu}$ = 50\% for $E_\mu~\textgreater~10$ GeV) 
is considered with collection efficiency 0.16\% . }
\label{fig:3gp16ceresvar}
\end{figure}

We then probe further with 0.16\% collection. Figure~\ref{fig:3gp16ceresvar} shows the results at 0.16\% 
collection for showers at various energy of the primary in the range $10^{16}$ eV -- $10^{19}$ eV and 
realistic detector energy resolution.

\subsection{Hadron models}
\begin{figure}[h]
\centering
\includegraphics[width=9cm]{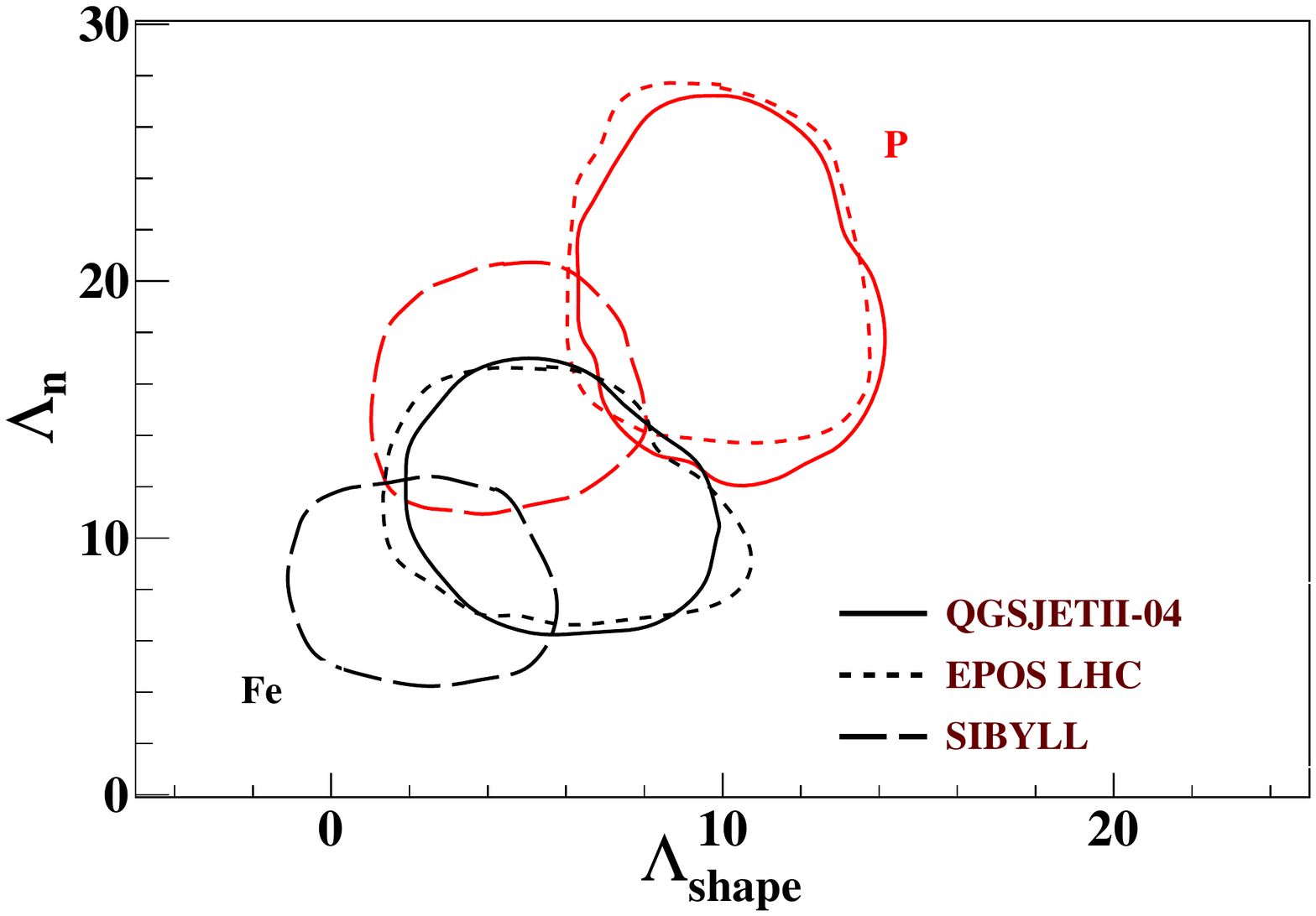}
\caption{Contours at 90\% confidence level in $\Lambda_{shape}$ -- $\Lambda_{n}$ plane for P and Fe 
primaries, for showers generated with three high energy interaction models QGSJETII-04, EPOS LHC 
and SIBYLL. Here, $E_p$ = $10^{16}$ eV and $\theta_p~=~0^{\circ}~\pm~2.5^{\circ}$.
An array of 
2m $\times$ 2m muon detectors ($\sigma_{E_\mu}$ = 20\% for $E_\mu~\textless~10$ GeV, $\sigma_{E_\mu}$ = 50\% for $E_\mu~\textgreater~10$ GeV) 
is considered with collection efficiency 0.16\%.}
\label{fig:HadModel}
\end{figure}

The very high energy hadronic interactions are not well understood yet and are a major source of uncertainty in our analysis 
as well as other analyses in the field. 
The previous results use the interaction model QGSJET II--04. Here we probe the fluctuations arising from three different 
hadron interaction models: QGSJET II--04, EPOS LHC and SIBYLL, used to simulate the high energy interactions. Figure~\ref{fig:HadModel} compares 
 the 90\% confidence level contours for these three interaction models for air showers generated from 
P and Fe primaries. It can be easily seen that SIBYLL predicted shapes and numbers are significantly away from the other 
two models which in turn agree quite well with each other.

\newpage
\section{Summary and future prospects}
\label{sect:summary}

 In this study, we performed a simulation study using CORSIKA to combine the longitudinal profile of 
the muons, characterized by their energy spectrum and lateral spread,  with the depth at shower 
maximum ($X_{max}$) of an EAS initiated by a primary at ultra high energies ($10^{16}$ eV -- $10^{19}$ eV). 
Firstly, using proton and iron as the shower primaries, we show that the muon observables and $X_{max}$ 
together can be used to identify the primary with relatively high acceptance and confidence. 
We find that for 0.16\% muon collection using a detector 
array of 2m $\times$ 2m detectors 50m apart from each other, we can distinguish P and Fe primaries 
at an acceptance above 50\%. This study is then generalized with three groups of primaries 
light, medium and heavy. The hadronic interaction models at higher energies are a major source 
of uncertainties, and we find a significant deviation among the models tested in this study. This 
deviation can also be a tool to probe the models at the UHE region using cosmic air shower data. 

The simulation involved 2m $\times$ 2m detector stations for muon detection. We find that a muon 
detector which is more sensitive to lower energies ($E_{\mu} \textless 10$ GeV) improves the results. 
We will extend this study to detailed analysis of different interaction models.
We plan to build a prototype surface station of such dimensions based on a scalable and economic technology,
which would be sensitive to 
individual muons events and will be able to measure their energies and direction.

\section{Acknowledgements}
\label{acknow} 
The authors thank the Weizmann Institute and Yeda-Sela foundation for the generous support. 
This work was supported by a Pazy-Vatat grant for young scientists.
RB is the incumbent of the Arye and Ido Dissentshik Career Development Chair. The authors 
also thank Dr. Hagar Landsman, Dr. Lorne Levinson,  
Dr. Alessandro Manfredini and Mr. Nadav Priel for the discussions in the course of this work.

\bibliographystyle{JHEP}
\bibliography{CRreferences} 

\end{document}